\documentclass[12pt]{iopart}
\usepackage{iopams}

\input epsf.sty

\newlength{\TZ}
\newcommand{\BEQ}{\begin{equation}}     
\newcommand{\BEA}{\begin{eqnarray}}
\newcommand{\BD}{\begin{displaymath}}
\newcommand{\EEQ}{\end{equation}}       
\newcommand{\EEA}{\end{eqnarray}}
\newcommand{\ED}{\end{displaymath}}
\newcommand{\eps}{\varepsilon}          
\newcommand{\vph}{\varphi}              
\newcommand{\D}{{\rm d}}                

\newcommand{\expect}[1]{\left\langle #1 \right\rangle}
\newcommand{\wit}[1]{\widetilde{#1}}    

\renewcommand{\vec}[1]{\boldsymbol{#1}} 

\catcode`\@=11
\def\numberbysection{\@addtoreset{equation}{section}
        \def\theequation{\thesection.\arabic{equation}}}

\begin{document}

\title[Ageing \& scaling without detailed balance]{Ageing, dynamical scaling 
and its extensions in many-particle systems without detailed balance}

\author{Malte Henkel$^{1,2}$}
\address{$^1$Laboratoire de Physique des 
Mat\'eriaux,\footnote{Laboratoire associ\'e au CNRS UMR 7556} 
Universit\'e Henri Poincar\'e Nancy I, \\ 
B.P. 239, F -- 54506 Vand{\oe}uvre l\`es Nancy Cedex, France\footnote{permanent address}}
\address{$^2$Dipartamento di Fisica/INFN, Universit\`a di Firenze,
I - 50019 Sesto Fiorentino, Italy}

\begin{abstract}
Recent studies on the phenomenology of ageing in certain many-particle systems 
which are at a critical point of their non-equilibrium steady-states, are reviewed. Examples include the contact process, the
parity-conserving branching-annihilating random walk, two exactly
solvable particle-reaction models and kinetic growth models. 
While the generic scaling descriptions known
from magnetic system can be taken over, some of the scaling relations between
the ageing exponents are no longer valid. In particular, there is no
obvious generalization of the universal limit fluctuation-dissipation ratio. 
The form of the scaling function of the two-time response function is
compared with the prediction of the theory of local scale-invariance. 
\end{abstract}

\submitto{\JPCM}
\maketitle

\setcounter{footnote}{0}

\section{Introduction}

Symmetry principles have been extremely useful for the understanding of complex
many-body systems, where the interactions between the degrees of freedom are sufficiently strong as to render perturbative methods inapplicables. Here
we are interested in the slow non-equilibrium dynamics shown by many-body
systems which are rapidly brought out from some initial state (`quenched') 
to a region in phase-space where either the equilibrium state naturally 
generates a slow dynamics (this is for example realized for systems {\em at}
a critical point) or else into a coexistence region dominated by several
equivalent stationary states. One of the essential features of such systems
is that their properties depend on their `age', that is the time elapsed
since the quench. Of course, any biological system ages, but there is also 
`physical ageing' which arises even if the underlying microscopic dynamics is completely reversible. One might formally define {\em ageing} by 
this breaking of time-translation invariance, associated with a slow 
dynamics which generically leads to some form of dynamical scaling. 
Physical ageing was originally seen to occur in glassy 
systems \cite{Stru78} and has been used since prehistoric times by engineers in 
the processing of materials. Quite recently, it has been realized that 
very similar phenomena can also be found in simple magnets, without disorder 
nor frustrations. The study of these supposedly simpler systems may lead
to conceptual insights which in turn could become also fruitful in more
complex systems. The topic has been under intensive study, see
\cite{Bray94,Cugl02,Godr02,Cris03,Henk04,Kawa04,Cala05,Henk05,Gamb06,Henk06} 
for reviews. 
The reversibility of the microdynamics in systems undergoing physical ageing
means that their stationary states are equilibrium states. In numerical
simulations of such systems this is realized by choosing the
dynamics such as to satisfy detailed balance. 
In many systems undergoing physical ageing, detailed balance and consequently 
the relaxation towards {\em equilibrium} steady-states is taken for granted, as 
there are many textbook proofs \cite{vKam92,Zwan01} of detailed balance for closed, isolated systems. 

On the other hand, it has become increasingly clear from studies in anomalous chemical kinetics that several of the constitutive properties of ageing are naturally met in many situations. First, it is well-known that fluctuation 
effects may lead to slow, non-exponential relaxation in {\em irreversible} chemical reactions - not accounted for by mean-field schemes, see e.g.
\cite{Schu00,Henk03f} for reviews and references therein. Second, it was understood more recently through the work of Oshanin and collaborators 
\cite{Osha89a,Osha89b,Argy01,Copp04,Voit05a,Voit05b} that even for
{\em reversible} reactions a slow, non-exponential relaxation may generically
occur without the fine-tuning of parameters and furthermore, that the
steady-states to which relaxation occurs depend on the kinetic coefficients and hence cannot be equilibrium states. Consequently, detailed balance cannot be
valid in these systems. While in these studies long-range interactions (as they may naturally arise in reactions of large molecules or in studies of radiation
damage \cite{Argy01}) play an essential r\^ole, detailed balance need not be 
satisfied even in kinetic systems with contact interactions. For example,
in the system defined by the simultaneous reversible reactions 
$2A\longleftrightarrow \emptyset$, $2A\longleftrightarrow A$, 
$A\longleftrightarrow \emptyset$ with diffusive motion of single particles, 
detailed balance only holds is certain conditions on the reaction rates are
met \cite{Alca94}. Since detailed balance is already found to be broken in very simple reactions such as $2A\longleftrightarrow B$ \cite{Voit05a,Voit05b} or
$A+B\longrightarrow \emptyset$ \cite{Argy01,Copp04}, it is conceivable that the
phenomenon might be much more common. Furthermore, since sometime a slow
relaxation in kinetic models is brought into relationship with glassy dynamics
\cite{Maye06}, it is of interest to investigate to what extent the three 
essential properties of physical ageing -- slow dynamics, dynamical scaling and breaking of time-translation invariance -- are actually realized in chemical 
kinetics. 

Therefore, we shall review here recent progress about ageing in many-body 
systems with a more general dynamics where detailed balance is no longer 
required to hold and therefore non-equilibrium steady-states may arise. 
Since there are as yet only few studies available on these systems, comparison 
with ageing in simple magnets should be a useful guide. For a similar reason, 
we shall investigate the behaviour of models whose physical origin is very different which should lead to some insight about generic properties. 
Because of the possibility of non-equilibrium steady-states, 
the systems under consideration here are closer to
biological/chemical ageing than those considered up to now in studies 
of physical ageing. Remarkably, there is evidence that some dynamical 
symmetries recently discovered in physical ageing may also extend to this more
general class of systems.

\begin{figure}
\centerline{
\epsfysize=65mm
\epsffile{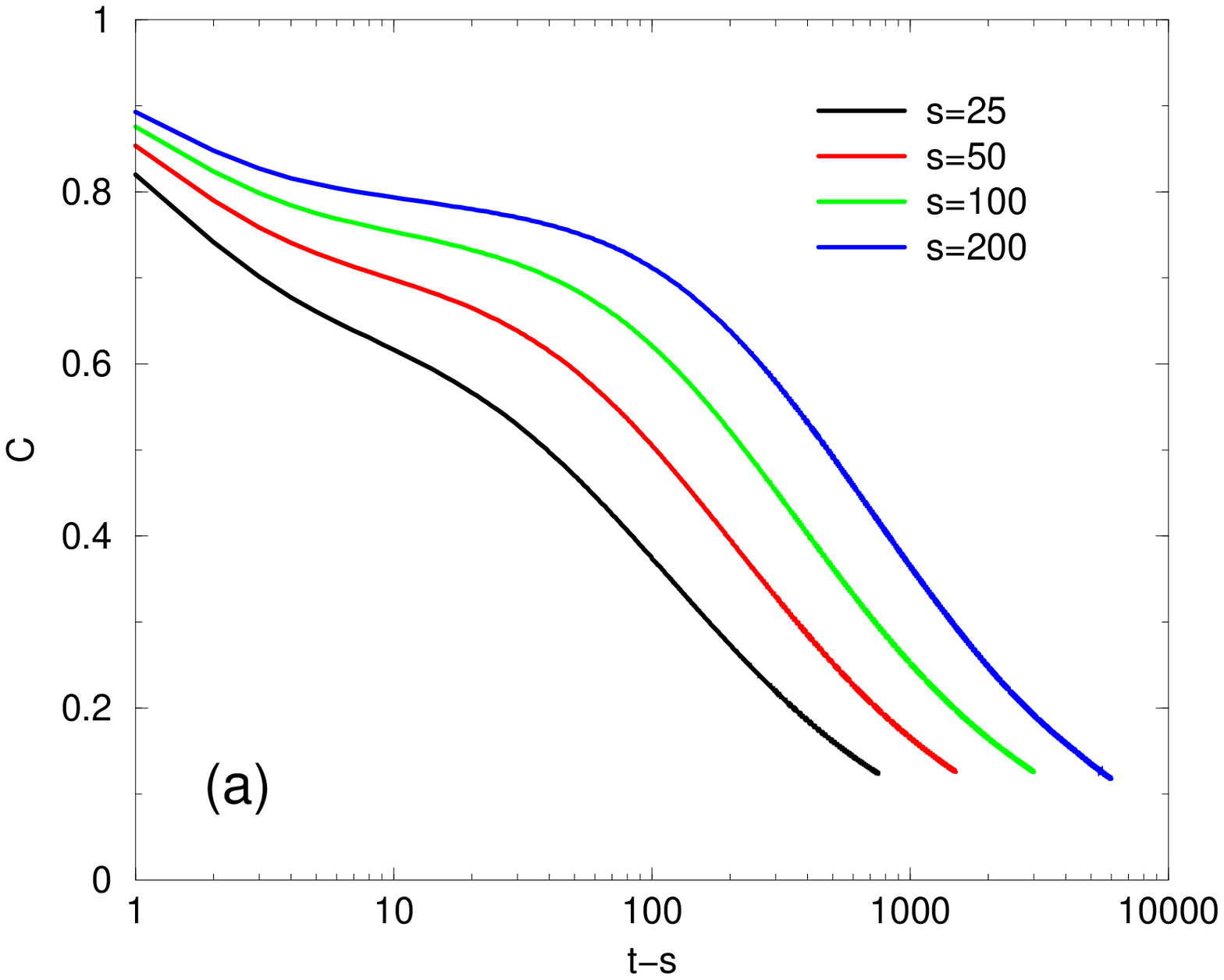}
\epsfysize=65mm
\epsffile{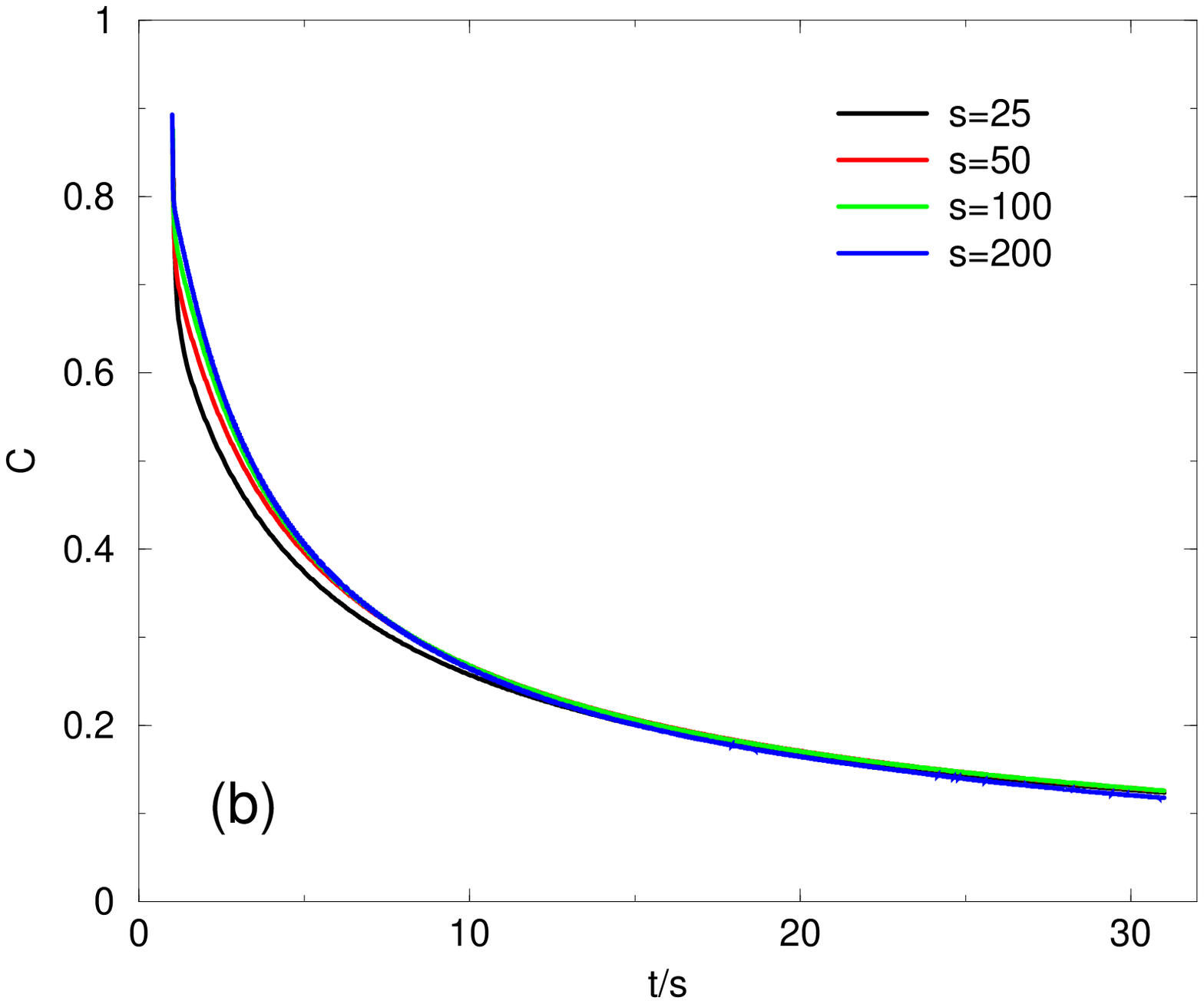}}
\caption[Scaling of 3D Ising model]{(a) Ageing and (b) dynamical 
scaling of the two-time autocorrelation
function $C(t,s)$ in the $3D$ Glauber-Ising model quenched to $T=3<T_c$, 
for several values of the waiting time $s$ \cite{Plei03}. 
\label{abb:Ising3d}
}
\end{figure}

We shall define the systems we want to study in the next section. Before
we come to that, however, we shall briefly recall for reference some of the
main results about the ageing of simple magnets. We assume throughout that the 
order-parameter is {\em non-conserved} by the dynamics and that the initial
state is totally disordered, unless explicitly stated otherwise. 
Besides the breaking
of time-translation invariance, ageing systems are often characterized by
dynamical scaling. It has become common to study ageing behaviour through 
the two-time autocorrelation and (linear) autoresponse functions
\BEA
C(t,s) &=& \langle \phi(t,\vec{r}) \phi(s,\vec{r}) \rangle ~~ 
\sim s^{-b} f_C(t/s) 
\label{gl:sC} \\
R(t,s) &=& \left.\frac{\delta\langle\phi(t,\vec{r})\rangle}{\delta 
h(s,\vec{r})}\right|_{h=0}
\sim s^{-1-a} f_{R}(t/s)
\label{gl:sR}
\EEA
where $\phi(t,\vec{r})$ is the order-parameter at time $t$ and location
$\vec{r}$ and $h(s,\vec{r})$ is the conjugate
magnetic field at time $s$ and location $\vec{r}$. 
The scaling behaviour is expected to apply
in the so-called {\em ageing regime} where $t,s\gg t_{\rm micro}$ 
and $t-s\gg t_{\rm micro}$, where $t_{\rm micro}$ is a microscopic time-scale.
We illustrate this in figure~\ref{abb:Ising3d} which shows the autocorrelation
function $C(t,s)$ of the $3D$ Ising model with (non-conserved) 
Glauber dynamics\footnote{Recall that the zero-temperature Glauber
model can be mapped, via a duality transformation \cite{Sigg77} or a similarity
transformation \cite{Sant97}, to the kinetic model $2A\longrightarrow\emptyset$ with single-particle diffusion.} \cite{Glau63} quenched 
from a fully disordered initial 
state to a temperature $T<T_c$. When plotting the data over against $t-s$, 
we see that the data depend on {\em both} $t-s$ and $s$, 
hence time-translation invariance is broken
and the system ages. Further,  with increasing 
values of the waiting time $s$, the system becomes `stiffer' and a plateau
close to the equilibrium value $C_{\rm eq}=M_{\rm eq}^2$ develops when
$t-s$ is not too large before the correlations fall off rapidly when
$t-s\to\infty$. When
replotting the same data over against $t/s$, a data collapse is found if $s$
is large enough which is evidence for dynamical scaling.

The distance of such systems from a global equilibrium state can be 
measured through the fluctuation-dissipation ratio, defined as \cite{Cugl94b}
\BEQ \label{gl:rfd}
X(t,s) := T R(t,s) \left( \frac{\partial C(t,s)}{\partial s}\right)^{-1}
\EEQ
At equilibrium, $X(t,s)=1$ from the fluctuation-dissipation theorem. One often
considers the limit fluctuation-dissipation ratio 
$X_{\infty} := \lim_{s\to\infty} \left( \lim_{t\to\infty} X(t,s)\right)$.\footnote{The order of the limits is crucial, since
$\lim_{t\to\infty}\left(\lim_{s\to\infty} X(t,s)\right)=1$.} 
For quenches to below $T_c$, one usually has $X_{\infty}=0$ but for
{\em critical} quenches onto $T=T_c$, it has been argued by Godr\`eche and Luck 
that $X_{\infty}$ should be a {\em universal} number \cite{Godr00b}, 
since it can be written as a ratio of two scaling amplitudes. 
This universality has been thoroughly confirmed for systems relaxing towards
equilibrium steady-states, see \cite{Cris03,Cala05} for recent reviews.  

\begin{figure}[t]
\centerline{
\epsfysize=55mm
\epsffile{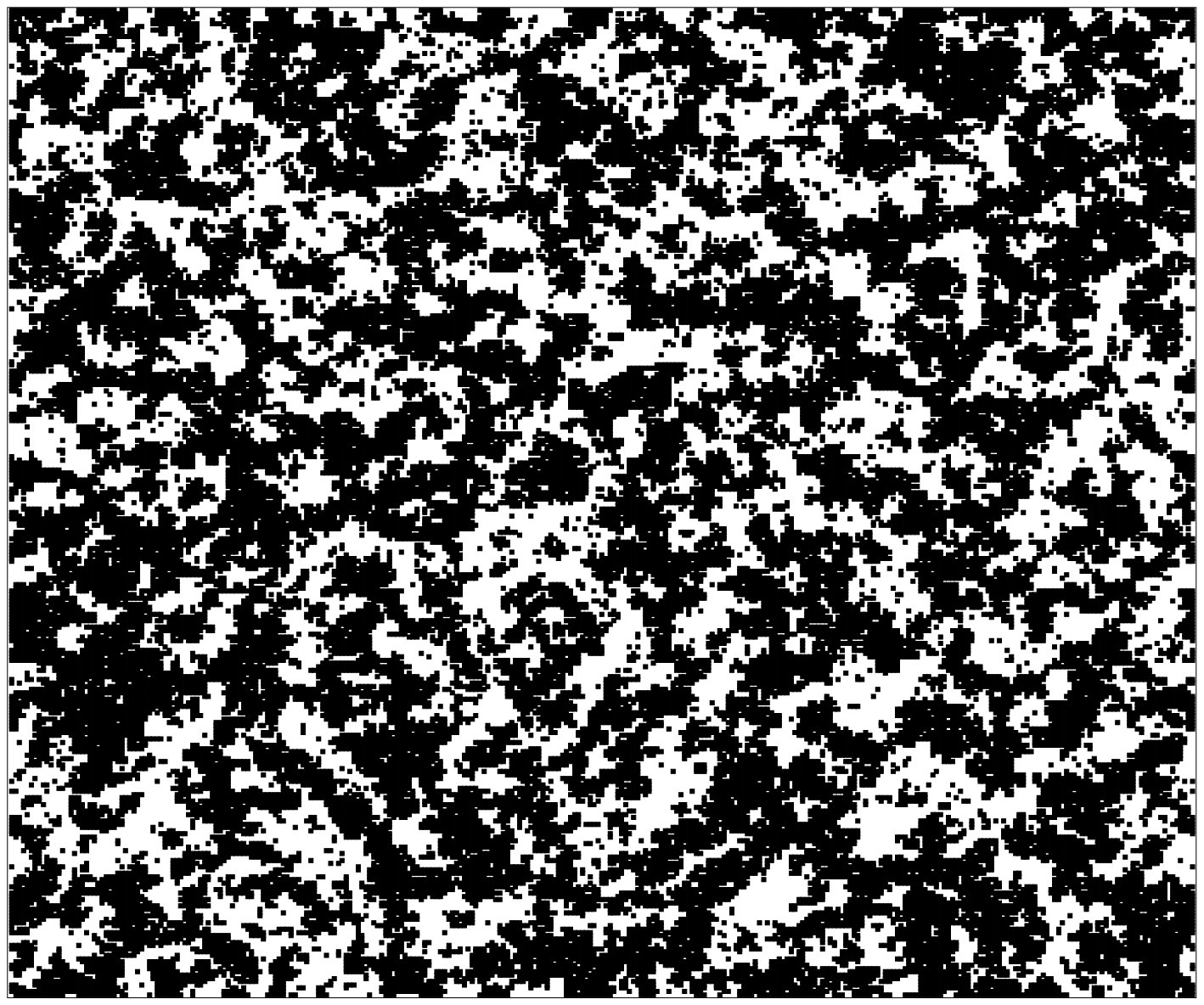} ~~~~~~
\epsfysize=55mm
\epsffile{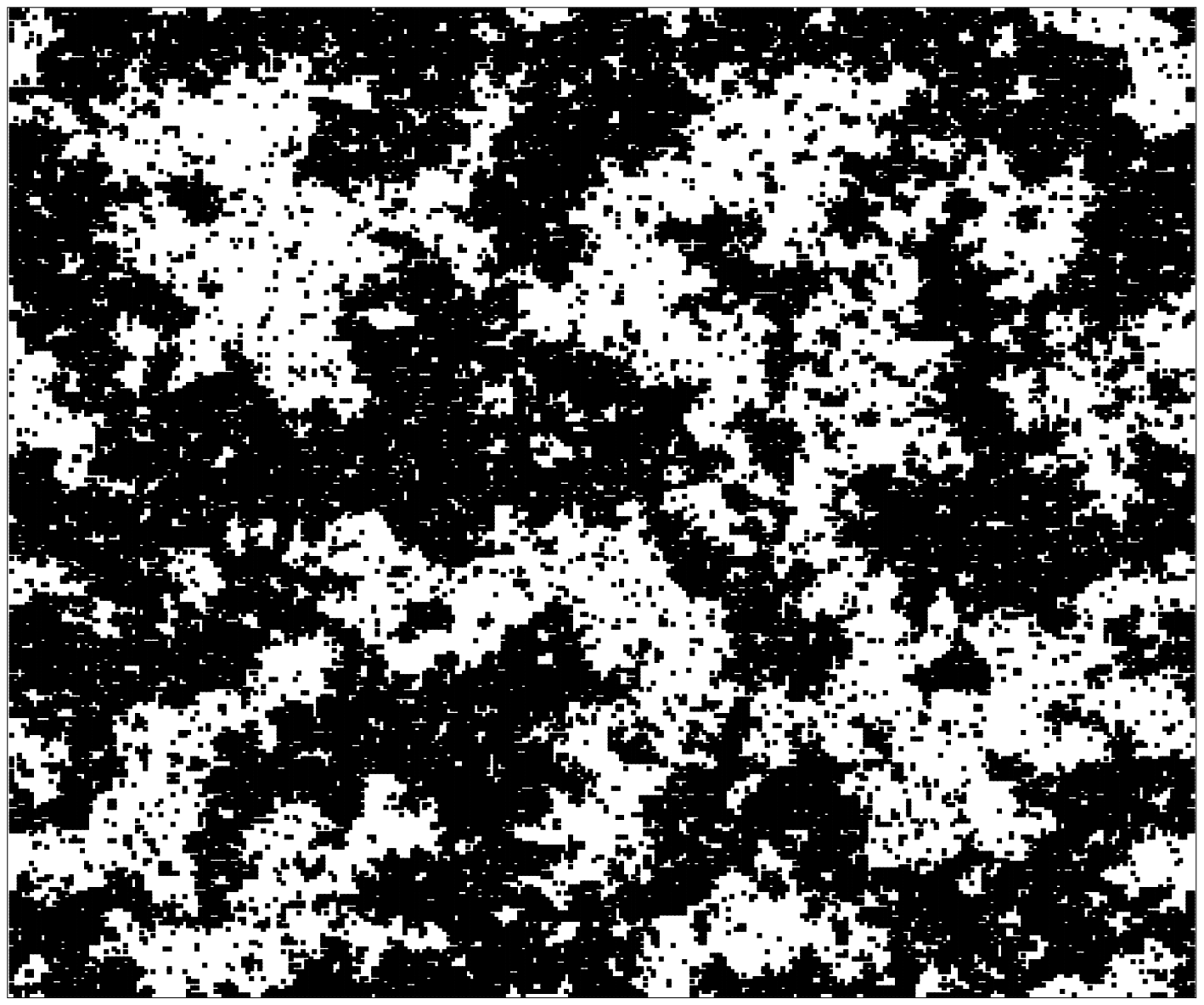}}
\caption[Growth of critical clusters]{Snapshots from the $2D$ Glauber-Ising
model quenched to $T=T_c$ from a disordered initial state at (left panel) 
$t=25$ and (right panel) $t=275$ MC time steps after the quench \cite{Plei03}. 
\label{abb:amasc}
}
\end{figure}

Furthermore, in writing eqs.~(\ref{gl:sC},\ref{gl:sR}) 
it was tacitly assumed that the scaling derives from the 
algebraic time-dependence of a single characteristic length-scale 
$L(t)\sim t^{1/z}$ which measures the linear size of correlated or ordered clusters and where $z$ is the dynamic exponent. For a $2D$ Glauber-Ising
model quenched to $T=T_c$ the growth of correlated clusters in illustrated in figure~\ref{abb:amasc} where the black/white site represent the two states
of the Ising spins. Then the above forms define the 
non-equilibrium exponents $a$ and $b$ and the scaling functions $f_C(y)$ and
$f_R(y)$. For large arguments $y\to \infty$, one generically expects
\BEQ
f_C(y) \sim y^{-\lambda_C/z} \;\; , \;\;
f_R(y) \sim y^{-\lambda_R/z}
\EEQ
where $\lambda_C$ and $\lambda_R$, respectively, are known as autocorrelation
\cite{Fish88,Huse89} and autoresponse exponents \cite{Pico02}. While
in non-disordered magnets with short-ranged initial conditions one usually
has $\lambda_C=\lambda_R$, this is not necessarily so if either
of these conditions is relaxed. From a field-theoretical point of view it 
is known that for a non-conserved order-parameter 
the calculation of $\lambda_{C,R}$ requires an independent renormalization and hence one cannot expect to find a scaling relation between
these and equilibrium exponents (including $z$) \cite{Jans89}. 
On the other hand, the values of the exponents $a$ and $b$ are known. For
quenches to $T=T_c$, the relevant length-scale is set by the time-dependent
correlation length $L(t)\sim \xi(t) \sim t^{1/z}$ and this leads to
$a=b=(d-2+\eta)/z$, where $\eta$ is a standard equilibrium exponent. For 
quenches into the ordered phase $T<T_c$, one usually observes simple scaling
of $C(t,s)=f_C(t/s)$, hence $b=0$.\footnote{This needs no longer be the case when
the ageing close to a free surface is considered \cite{Baum06a}.} 
The value of $a$ depends on whether the equilibrium
correlator is short- or long-ranged, respectively. These may be referred to as classes S and L, respectively and 
one has, see e.g. \cite{Cate00,Henk02a,Henk03e}
\BEQ
\hspace{-1truecm} C_{\rm eq}(\vec{r})\sim \left\{\begin{array}{l} e^{-|\vec{r}|/\xi} \\
|\vec{r}|^{-(d-2+\eta)} \end{array} \right. 
\;\; \Longrightarrow \;\; 
\left\{\begin{array}{l} \mbox{\rm class S} \\ \mbox{\rm class L}\end{array}\right.
\;\; \Longrightarrow \;\; 
a = \left\{ \begin{array}{c} 1/z \\ (d-2+\eta)/z \end{array} \right.
\EEQ
Examples for short-ranged models (class S) include the Ising or Potts 
models in $d>1$
dimensions (and $T<T_c$), while all systems quenched to criticality, or
the spherical model or the $2D$ XY model below the Kosterlitz-Thouless transition are examples for long-ranged systems (class L).

In equilibrium critical phenomena, it is well-known that the standard scale-invariance can, under quite weak conditions, be extended to
a {\em conformal} invariance. Roughly, a conformal transformation is
a scale-transformation $\vec{r}\mapsto b \vec{r}$ with a space-dependent 
rescaling factor $b=b(\vec{r})$ (such that angles are kept unchanged).  
In particular, in two dimensions conformal invariance allows to derive
from the representation theory of the conformal (Virasoro) algebra the possible values of the critical exponents, to set up a list of possible universality classes, calculate explicitly all $n$-point correlation functions and so on
\cite{Bela84,Card90}. 
One might wonder whether a similar extension might be possible 
at least in some instances of dynamical scaling and further ask {\it whether
response functions or correlation functions might be found from their
covariance under some generalized dynamical scaling with a space-time-dependent
rescaling factor} $b=b(t,\vec{r})$ \cite{Henk94,Henk02}~?  We shall discuss the 
question here in the specific context of ageing and shall focus on what
can be said about the scaling functions $f_{C,R}(y)$ in a model-independent way. 

A useful starting point is to consider the symmetries of the free diffusion 
(or free Schr\"odinger) equation
\BEQ \label{gl:diffu}
2{\cal M} \partial_t \phi = \Delta \phi
\EEQ
where $\Delta=\vec{\nabla}\cdot\vec{\nabla}$ is the spatial laplacian and 
the `mass' $\cal M$ can be seen as a kinetic coefficient. Indeed, 
it was already shown by Lie more than a century ago that this equation
has more symmetries than the trivial translation- and rotation-invariances. 
Consider the so-called {\em Schr\"odinger-group} defined through the
space-time transformations
\BEQ \label{gl:5:2:Schr}
t \mapsto t' = \frac{\alpha t+\beta}{\gamma t+\delta} \;\; ; \;\;
\vec{r} \mapsto \vec{r}' = 
\frac{\matrix{R}\vec{r} + \vec{v}t + \vec{a}}{\gamma t+\delta} 
\;\; , \;\;
\alpha\delta - \beta\gamma =1
\EEQ
where $\alpha,\beta,\gamma,\delta,\vec{v},\vec{a}$ are real (vector) 
parameters and $\matrix{R}$ is a rotation matrix in $d$ spatial dimensions. 
The group acts projectively on a solution $\phi$ of the diffusion 
equation through 
$(t,\vec{r})\mapsto g(t,\vec{r})$, $\phi\mapsto T_g \phi$
\BEQ \label{gl:5:2:Schrpsi}
\left(T_g \phi\right)(t,\vec{r}) = 
f_g(g^{-1}(t,\vec{r}))\,\phi(g^{-1}(t,\vec{r}))
\EEQ
where $g$ is an element of the Schr\"odinger group and the companion function reads \cite{Nied72,Perr77}
\BEQ \label{gl:5:2:Schrf}
\hspace{-2.2truecm}
f_{g}(t,\vec{r}) = (\gamma t+\delta)^{-d/2} 
\exp\left[ -\frac{{\cal M}}{2} \frac{\gamma \vec{r}^2+
2\matrix{R}\vec{r}\cdot(\gamma\vec{a}-\delta\vec{v})+\gamma\vec{a}^2-
t\delta\vec{v}^2+2\gamma\vec{a}\cdot\vec{v}}{\gamma t+\delta}\right]
\EEQ
It is then natural to include also arbitrary phase-shifts of the wave function 
$\phi$ within the Schr\"odinger group {\sl Sch}($d$). In what follows, we denote 
by $\mathfrak{sch}_d$ the Lie algebra of {\sl Sch}($d$). The Schr\"odinger
group so defined is the largest group which maps {\em any} solution of the
free Schr\"odinger equation (with $\cal M$ fixed) onto another solution. 
This is easily seen in $d=1$ by introducing the Schr\"odinger operator
\BEQ
{\cal S} := 2 M_0 Y_{-1} - Y_{-1/2}^2
\EEQ
The Schr\"odinger Lie algebra
$\mathfrak{sch}_1=\langle X_{-1,0,1},Y_{-\frac{1}{2},\frac{1}{2}},M_0\rangle$
is spanned by the infinitesimal generators of temporal and spatial
translations ($X_{-1},Y_{-1/2}$), Galilei-transformations ($Y_{1/2}$),
phase shifts ($M_0$), space-time dilatations with $z=2$ ($X_0$) and so-called
special transformations ($X_1$). Explicitly, the generators read \cite{Henk94}
\BEA
X_n &=& -t^{n+1}\partial_t -\frac{n+1}{2} t^n r\partial_r -\frac{n(n+1)}{4}
{\cal M} t^{n-1} r^2 - \frac{x}{2}(n+1) t^n \nonumber \\
Y_m &=& -t^{m+1/2}\partial_r -\left( m+\frac{1}{2}\right) {\cal M} t^{m-1/2} r
\label{gl:5:2:SchrGen} \\
M_n &=& -{\cal M} t^n \nonumber
\EEA
Here $x$ is the scaling dimension and ${\cal M}$ is the {mass} of the
scaling operator $\phi$ on which these generators act. 
The non-vanishing commutation relations are
\BEA
\left[ X_n , X_{n'} \right] &=& (n-n') X_{n+n'} \;\; , \;\;
\left[ X_n , Y_m \right] \:=\: \left(\frac{n}{2}-m\right) Y_{n+m} 
\nonumber \\ 
\left[ X_n , M_{n'} \right] &=& -n' M_{n+n'} \;\; , \;\; 
\left[ Y_m , Y_{m'} \right] \:=\: (m-m') M_{m+m'} 
\label{gl:5:2:SchAlg}
\EEA
\setcounter{footnote}{1}
The invariance of the diffusion equation under the action of $\mathfrak{sch}_1$
is now seen from the following commutators which follow from the
explicit form (\ref{gl:5:2:SchrGen})
\BEA
\left[{\cal S}, X_{-1}\right] &=& \left[ {\cal S}, Y_{\pm 1/2} \right]
\:=\: \left[ {\cal S}, M_0\right] \:=\: 0 
\nonumber \\
\left[{\cal S}, X_0 \right] &=& -{\cal S}
\;\; , \;\; \qquad \;\;\:
\left[ {\cal S}, X_1 \right] \:=\: -2t {\cal S} - (2x-1) M_0 
\EEA
Therefore, {\em for any solution $\phi$ 
of the Schr\"odinger equation ${\cal S}\phi=0$ with
scaling dimension $x=1/2$, the infinitesimally transformed solution 
${\cal X}\phi$ with ${\cal X}\in\mathfrak{sch}_1$ also satisfies the 
Schr\"odinger equation ${\cal S}{\cal X}\phi=0$} \cite{Kast68,Nied72,Hage72}. 
For applications to ageing, we must consider to so-called {\em ageing algebra} 
$\mathfrak{age}_1 =\langle X_{0,1},Y_{-\frac{1}{2},\frac{1}{2}},M_0\rangle
\subset \mathfrak{sch}_1$ (without time-translations) 
which is a true subalgebra of $\mathfrak{sch}_1$. 
Extensions to $d>1$ are straightforward. 

What is the usefulness of knowing dynamical symmetries of free, simple
diffusion for the understanding of non-equilibrium kinetics ? 
One way of setting up the problem would be to write down a stochastic
Langevin equation for the order-parameter. The simplest case is usually
considered to be a dynamics without macroscopic conservation laws (model A), 
where one would have \cite{Hohe77} 
\BEQ \label{gl:5:Langevin}
2{\cal M} \frac{\partial\phi}{\partial t} = 
\Delta \phi - \frac{\delta {\cal V}[\phi]}{\delta \phi} +\eta
\EEQ
where $\cal V$ is the Ginzburg-Landau potential and $\eta$
is a gaussian noise which describes the coupling to an external heat-bath and
the initial distribution of $\phi$. At first sight, there appear to be no
non-trivial symmetries, because (\ref{gl:5:Langevin}) 
cannot be Galilei-invariant, because of the noise term $\eta$. 
To understand this physically, consider a magnet which
is a rest with respect to a homogeneous heat-bath at temperature $T$. 
If the magnet is moved with a constant velocity with respect to 
the heat-bath, the effective
temperature will now appear to be direction-dependent, and the heat-bath
is no longer homogeneous. However, this difficulty can be avoided as 
follows \cite{Pico04}: {\em split the Langevin equation into a
`deterministic' part with non-trivial symmetries and a `noise' part and then
show using these symmetries that all averages can be reduced exactly to
averages within the deterministic, noiseless theory}. Technically, one
first constructs in the standard fashion (Janssen-de Dominicis procedure)
\cite{deDo78,Jans92} the associated stochastic field-theory with action 
$J[\phi,\wit{\phi}]$ where $\wit{\phi}$ is the response field associated to 
the order-parameter $\phi$. Second, decompose the action into two parts
\BEQ \label{gl:5:JanDom}
J[\phi,\wit{\phi}] = J_0[\phi,\wit{\phi}] + J_b[\wit{\phi}]
\EEQ
where  
\BEQ \label{gl:5:JanDomdet}
J_0[\phi,\wit{\phi}]=\int_{\mathbb{R}_+\times\mathbb{R}^d}\!\D t\D\vec{r}\; 
\wit{\phi}\left(2{\cal M}\partial_t\phi-\Delta\phi+\frac{\delta{\cal V}}{\delta\phi}\right)
\EEQ 
contains the terms coming from the `deterministic' part of
the Langevin equation ($\cal V$ is the self-interacting `potential') whereas
\BEQ \label{gl:5:JanDombruit}
\hspace{-1truecm}
J_b[\wit{\phi}] = -T \int_{\mathbb{R}_+\times\mathbb{R}^d}\!\D t\D\vec{r}\: 
\wit{\phi}(t,\vec{r})^2 
-\frac{1}{2}
\int_{\mathbb{R}^{2d}} \!\D\vec{r}\,\D\vec{r}'\:  
{\wit{\phi}}(0,\vec{r})a(\vec{r}-\vec{r}'){\wit{\phi}}(0,\vec{r}')
\EEQ
contains the `noise'-terms coming from (\ref{gl:5:Langevin}) \cite{Jans92}. 
It was assumed here that $\langle \phi(0,\vec{r})\rangle =0$ and
$a(\vec{r})$ denotes the initial two-point correlator 
\BEQ
a(\vec{r}) := C(0,0;\vec{r}+\vec{r}',\vec{r}') = 
\langle \phi(0,\vec{r}+\vec{r}') \phi(0,\vec{r}')\rangle =a(-\vec{r})
\EEQ
while the last relation follows from spatial translation-invariance 
which we shall admit throughout.

It is instructive to consider briefly the case of a free field, where
${\cal V}=0$. Variation of (\ref{gl:5:JanDom}) with respect to $\wit{\phi}$ and
$\phi$, respectively, then leads to the equations of motion
\BEQ \label{gl:5:equamovi}
2{\cal M}\partial_t \phi = \Delta \phi + T \wit{\phi} \;\; , \;\;
-2{\cal M}\partial_t\wit{\phi} = \Delta\wit{\phi}
\EEQ
The first one of those might be viewed as a Langevin equation if $\wit{\phi}$ is
interpreted as a noise. Comparison of the two equations of motion 
(\ref{gl:5:equamovi}) shows that if the order-parameter $\phi$ is characterized by
the `mass'\index{mass} $\cal M$ (which by physical convention is positive), 
then the associated response field $\wit{\phi}$ is 
characterized by the {\em negative} mass $-{\cal M}$. 
This characterization remains valid beyond free fields. 

We now concentrate on actions $J_0[\phi,\wit{\phi}]$ which
are Galilei-invariant. This means that if $\langle .\rangle_0$ denotes the
averages calculated only with the action $J_0$, 
the Bargman superselection rules \cite{Barg54}
\BEQ \label{gl:5:Bargman}
\left\langle \underbrace{\phi\ldots\phi}_n \; \underbrace{\wit{\phi}\ldots
\wit{\phi}}_m \right\rangle_0 \sim \delta_{n,m}
\EEQ
hold true. It follows that both response and correlation functions can be
exactly expressed in terms of averages with respect to the deterministic part
alone. For example (we suppress
for notational simplicity the spatial coordinates) \cite{Pico04}
\BEQ
\hspace{-1truecm} 
R(t,s) = \left.\frac{\delta\langle \phi(t)\rangle}{\delta h(s)}\right|_{h=0}
= \left\langle \phi(t) \wit{\phi}(s) \right\rangle
= \left\langle \phi(t) \wit{\phi}(s)\, e^{-J_b[\wit{\phi}]} \right\rangle_0
= \left\langle \phi(t) \wit{\phi}(s) \right\rangle_0 
\label{gl:5:Rrauschlos}
\EEQ
where the `noise' part of the action was included in the
observable and the Bargman superselection rule
(\ref{gl:5:Bargman}) was used. In other words, 
{\em the two-time response function
does not depend explicitly on the `noise' at all}. The correlation function
is reduced similarly 
\BEA
\lefteqn{ 
C(t,s;\vec{r}) = 
T \int_{\mathbb{R}_+\times\mathbb{R}^d}\!\D u\D\vec{R}\: 
\left\langle \phi(t,\vec{r}+\vec{y})\phi(s,\vec{y})\wit{\phi}(u,\vec{R})^2
\right\rangle_0 
}
\nonumber \\
& & +\frac{1}{2} \int_{\mathbb{R}^{2d}}\!\D\vec{R}\D\vec{R}'\: 
a(\vec{R}-\vec{R}')
\left\langle \phi(t,\vec{r}+\vec{y})\phi(s,\vec{y})
\wit{\phi}(0,\vec{R})\wit{\phi}(0,\vec{R}') \right\rangle_0
\label{gl:5:Crauschlos} 
\EEA
Only terms which depend
explicitly on the `noise' remain -- recall the vanishing of 
the `noiseless' two-point function 
$\langle \phi(t)\phi(s)\rangle_0=0$ because of the Bargman superselection rule. 

Therefore, the dynamical symmetries of non-equilibrium
kinetics are characterized by the `deterministic' part of Langevin equation.
Such deterministic non-linear diffusion/Schr\"odinger equations with 
$\mathfrak{age}_1$ or $\mathfrak{sch}_1$ as a dynamical symmetry can be 
explicitly constructed \cite{Stoi05} but we shall not go into the details
here. Since all quantities of interest will reduce to some kind of 
response function,
one may calculate them from the {\it requirement that they transform covariantly
under the action ageing subgroup} (with Lie algebra $\mathfrak{age}_d$) 
obtained from the Schr\"odinger group when leaving out
time-translations. In this survey, we shall concentrate on the
two-time autoresponse function $R(t,s)$ for which 
the requirement of covariance reduces to the two conditions 
$X_0 R(t,s) = X_1 R(t,s)=0$. Since time-translations are not
included in the ageing group, the generators $X_n$ can be
generalized from (\ref{gl:5:2:SchrGen}) to the following form
\BEQ \label{gl:Xnext}
\hspace{-2truecm}
X_n = -t^{n+1}\partial_t - \frac{n+1}{2} t^n r\partial_r
-\frac{(n+1)n}{4}{\cal M}t^{n-1} r^2 - \frac{x}{2}(n+1)t^n -\xi n t^n
\;\; ; \;\; n\geq 0
\EEQ
where $\xi$ is a new quantum number associated with the field $\phi$ on which the generators $X_n$ act. This last term can only be present for systems
out of an equilibrium state (the requirement of time-translation invariance
and $[X_1,X_{-1}]=2X_0$ lead to $\xi=0$). Solving the two differential
equations for $R$ gives the explicit form of $R(t,s)$, see (\ref{gl:Rf}) below. 

While this discussion was carried out explicitly for the case $z=2$, it is
tempting to try and generalize this idea to more general values of $z$. 
In this way, the notion of {\em local scale-transformation} has been
introduced, which is based on the following main assumptions \cite{Henk02}. 
\begin{enumerate}
\item In principle, the following conformal time-transformations should be
included 
\BEQ \label{gl:5:3:Moeb}
t\mapsto t' = \frac{\alpha t + \beta}{\gamma t +\delta} \;\; ; \;\;
\alpha\delta - \beta\gamma =1
\EEQ
For applications to ageing, however, time-translations generated by $\beta$
must be left out (generalizing the restriction $\mathfrak{sch}_d\to\mathfrak{age}_d$). 
\item The generator $X_0$ of scale-transformations is
\BEQ \label{gl:5:3:X0gen}
X_0 = - t \partial_t - \frac{1}{z} r \partial_r - \frac{x}{z}
\EEQ
where $x$ is the scaling dimension of the quasi-primary operator on which
$X_0$ is supposed to act. Physically, this implies that there is a single 
relevant length scale $L(t)\sim t^{1/z}$. 
\item Spatial translation-invariance is required. 
\end{enumerate}
Generators for infinitesimal local scale-transformations have been explicitly
constructed and
it can be shown that for any value of $z$ there is a linear invariant
equation, analogous to (\ref{gl:diffu}) \cite{Henk02}. 
{\it Local scale-invariance} (LSI) assumes in particular that the two-time 
response functions transform covariantly under these local 
scale-transformations, hence $X_0R=X_1R=0$. This leads to the 
prediction \cite{Henk02,Pico04,Henk05a,Henk06a}
\BEQ \label{gl:Rf}
\hspace{-1truecm}
R(t,s) = \left\langle \phi(t)\wit{\phi}(s)\right\rangle = s^{-1-a} f_R(t,s) 
\;\; , \;\; 
f_R(y) = f_0\, y^{1+a'-\lambda_R/z} (y-1)^{-1-a'}
\EEQ
where the exponents $a,a',\lambda_R/z$ are related to $x,\xi,\wit{x},\wit{\xi}$ 
and $f_0$ is a normalization constant.\footnote{We point out that the
prediction (\ref{gl:Rf}) as well as the explicit form  (\ref{gl:Xnext}) of
$X_n$, valid for $z=2$, 
assume that the mean order-parameter $\langle \phi(0,\vec{r})\rangle=m_0=0$
at the initial moment when the quench to $T<T_c$ or $T=T_c$ is made.}

\begin{table}
\caption[Tabelle1]{Magnetic systems quenched into the coexistence phase 
($T<T_c$) which satisfy (\ref{gl:Rf}) with the exponents
$a=a'$ and $\lambda_R$. $d$ is the spatial dimension and the 
numbers in brackets estimate the numerical uncertainty in the last digit(s). 
In the spherical model, long-range initial conditions are included and
in the long-range spherical model the exchange couplings decay as
$J_{\vec{r}}\sim |\vec{r}|^{-d-\sigma}$. In the bond-disordered Ising model,
the couplings are taken homogeneously from the interval $[1-\eps/2,1+\eps/2]$.
Then $z=z(T,\eps)=2+\eps/T$ exactly \cite{Paul04,Paul05} and one observes roughly $1.3\lesssim \lambda_R(T,\eps) \lesssim 1.7$.
\label{Tabelle1}
}
\begin{center}
\begin{tabular}{||l|ll|rr|c|l||} \hline\hline\hline
model & $d$ & $z$ & $a=a'$ & $\lambda_R$ & & Ref. \\ \hline\hline
Ising & 2 & 2 & $1/2$      & 1.26(1) & & \cite{Henk03b} \\
      & 2 & 2 & $\simeq 0.5$ & 1.24(2) & & \cite{Lore06,Jank06} \\
      & 3 & 2 & $1/2$      & 1.60(2) & & \cite{Henk03b} \\ \hline
Potts-3 & 2 & 2 & 0.49     & 1.19(3) & & \cite{Lore06,Jank06} \\\hline
Potts-8 & 2 & 2 & 0.51     & 1.25(1) & & \cite{Lore06,Jank06} \\\hline
XY      & 3 & 2 & 0.5      & 1.7     & & \cite{Abri04b} \\
XY spin wave & $\geq 2$ & 2 & $d/2-1$ & $d$ & angular response & \cite{Pico04} \\\hline
spherical & $>2$ & 2 & $d/2-1$ & $(d-\alpha)/2$ & $C_{\rm ini}(\vec{r})\sim |\vec{r}|^{-d-\alpha}$ & \cite{Newm90,Pico02} \\\hline
long-range  & $>2$ & $\sigma$ & $d/\sigma-1$ & $d/2$ & $0<\sigma<2$ & \\
spherical & $\leq 2$ & $\sigma$ & $d/\sigma-1$ & $d/2$ & $0<\sigma<d$ & \cite{Cann01} \\\hline\hline
diluted Ising & 2 & $2+\eps/T$ & $1/z(T,\eps)$ & $\lambda_R(T,\eps)$ & disordered & \cite{Henk06b}
\\\hline\hline\hline
\end{tabular}\end{center}
\end{table}

\begin{table}
\caption[Tabelle 2]{Systems quenched to a critical point of their
stationary state which satisfy (\ref{gl:Rf}) with the exponents $a$, $a'$ and $\lambda_R/z$. $d$ is the spatial dimension and the 
numbers in brackets estimate the uncertainty in the last digit(s). 
{\sc csm} stands for the spherical model with a conserved order-parameter,
{\sc fa} denotes the Frederikson-Andersen model, {\sc nekim} is the
non-equilibrium kinetic Ising model and {\sc bcp} and {\sc bpcp} denote the
bosonic contact and pair-contact processes 
(see eqs. (\ref{gl:4:bcp:Krit},\ref{gl:def_alphaC}) for the
definitions of the control parameter $\alpha$ and of $\alpha_C$), respectively. 
In the spherical model, long-range initial correlations 
$C_{\rm ini}(\vec{r})\sim |\vec{r}|^{-d-\alpha}$ were considered. 
If $d+\alpha>2$, these reduce to short-ranged initial correlations 
(denoted {\sc s}), but for $d+\alpha<2$ a new class {\sc l} arises. 
In those models described by a Langevin equation, one has used throughout, with 
the exception of the {\sc csm}, the simple white
noise $\langle\eta(t,\vec{r})\eta(s,\vec{r}')\rangle
=2T\delta(\vec{r}-\vec{r}')\delta(t-s)$.
\label{Tabelle2}
}
\begin{center}
\begin{tabular}{||l|r|rrr|cl||} \hline\hline\hline
model & \multicolumn{1}{c|}{$d$} & $a$ & $a'-a$  & $\lambda_R/z$ & & Ref.  \\\hline\hline
random walk & & -1 & 0 & 0 & & \cite{Cugl94b} \\ \hline
OJK-model & & $(d-1)/2$ & $-1/2$  & $d/4$ &  & \cite{Bert99,Maze04,Henk05a}\\\hline
Ising & 1 & 0 & $-1/2$  & $1/2$ &  & \cite{Godr00a,Lipp00,Henk03d}\\ 
      & 2 & $0.115$ & $-0.187(20)$ & $0.732(5)$ & & \cite{Plei05,Henk06a}\\ 
      & 3 & $0.506$ & $-0.022(5)$  & $1.36$     & &\cite{Plei05,Henk06a}\\\hline
XY    & 3 & 0.52    & 0            & 1.34(5)    & & \cite{Abri04b} \\\hline
spherical $d>2$ & $<4$ & $d/2-1$ & 0 & $d/2-\alpha/4-1/2$ & {\sc l} & \cite{Pico02} \\
          & $>4$ & $d/2-1$ & 0 & $(d-\alpha)/4 +1/2$ & {\sc l} &
            \cite{Pico02}\\
          & $<4$ & $d/2-1$ & 0 & $3d/4-1$ & {\sc s} & \cite{Godr00b} \\
	  & $>4$ & $d/2-1$ & 0 & $d/2$ & {\sc s} & \cite{Godr00b} \\\hline\hline
{\sc csm} & $>2$ & $d/4-1$ & 0 & $(d+2)/4$ &   & \cite{Baum06e} \\ \hline\hline
disordered Ising     & $4-\eps$ & $1-\frac{1}{2}\sqrt{\frac{6\eps}{53}}$ & 0 & $3-\frac{1}{2}\sqrt{\frac{6\eps}{53}}$ & O($\eps$), $\log$ & \cite{Cala02a,Sche05,Sche06}\\\hline\hline
{\sc fa} & $>2$ & $1+d/2$ & $-2$  & $2+d/2$ &  & \cite{Maye06} \\
         & $1$ & $1$ & $-3/2$ & $2$ &  & \cite{Maye06,Maye04}\\\hline
Ising spin glass & 3 & $0.060(4)$ & $-0.76(3)$ & $0.38(2)$ &  & \cite{Henk04a,Henk05a} \\ \hline\hline 
contact process & 1 & $-0.681$ & $+0.270(10)$ & $1.76(5)$ & $t/s\gtrsim 1.1$ & \cite{Enss04,Hinr06,Henk06a} \\
 & $>4$ & $d/2-1$ & 0 & $d/2+2$ & & \cite{Rama04} \\\hline
{\sc nekim} & 1 & -0.430(4) & 0 & 1.9(1) & & \cite{Odor06} \\\hline
{\sc bcp} & $\geq 1$ & $d/2-1$ & 0 & $d/2$ & & \cite{Baum05a} \\\hline
{\sc bpcp} & $>2$ & $d/2-1$ & 0 & $d/2$ & $\alpha\leq\alpha_C$ & \cite{Baum05a} \\\hline\hline\hline 
\end{tabular}\end{center}
\end{table}

Starting with \cite{Henk01}, the prediction (\ref{gl:Rf}) has been reproduced 
in many different spin systems
and we list examples quenched to below criticality in table~\ref{Tabelle1}
and quenched to the critical point in table~\ref{Tabelle2}. For $T<T_c$,
it is found empirically that $a=a'$ in all examples considered so far. 
We point out that agreement with local scale-invariance eq.~(\ref{gl:Rf}) is not
only obtained for systems where the dynamical exponent is $z=2$, but that
rather there exist quite a few examples where $z$ can become 
considerably larger or smaller than 2. It must be remembered, 
however, that the above 
derivation of (\ref{gl:Rf}) for a stochastic Langevin equation has for the time 
being only been carried out for $z=2$\footnote{See section~4 for a recent extension of the method to $z=4$.} and the justification of
$X_0R=X_1R=0$ remains an open problem for $z\ne 2$ although the result
(\ref{gl:Rf}) seems to work remarkably well.  
Still, it is non-trivial that a relatively simple
extension of dynamical scaling should be capable of making predictions
which can be reproduced in physically quite different systems. 

A few comments are still needed: (i) for the XY model in the 
spin-wave approximation (table~\ref{Tabelle1}), 
eq.~(\ref{gl:Rf}) holds for the response of the 
angular variable $\phi=\phi(t,\vec{r})$ which is related to the XY spin through
$\vec{S}=(\cos \phi,\sin \phi)$. Magnetic responses have a different scaling
form \cite{Bert01,Abri04a}. (ii) in the critical disordered Ising model
(table~\ref{Tabelle2}) one finds a logarithmic scaling form
$R(t,s)=(r_0+r_1\ln (t-s))f_R(t/s)$ \cite{Cala02a,Sche05,Sche06} such that
the computed $f_R(y)$ is consistent with (\ref{gl:Rf}) to one-loop
order, or up to terms of order ${\rm O}(\eps)$. (iii) Finally, a two-loop
calculation of the critical non-conserved O($n$)-model does produce in 
$4-\eps$ dimensions an expression for $f_R(y)$ which is incompatible
with (\ref{gl:Rf}) \cite{Cala02} and a similar result is anticipated in
$2+\eps$ dimensions \cite{Fedo06}; although the one-loop results are still
compatible \cite{Cala01,Cala02,Fedo06}. Should one conclude from these studies 
that for $T=T_c$ the prediction (\ref{gl:Rf}) and by implication local 
scale-invariance can only hold approximatively~? 
This might well be a subtle question. Deviations
between (\ref{gl:Rf}) and the field-theoretical studies typically arise when 
$t/s\approx 1$. However, in this region the field-theoretical 
results for $f_R(y)$ do not
agree with the ones of non-perturbative numerical studies \cite{Plei05}. 
Since the perturbative expansion usually carried out in field-theoretical
studies does not necessarily take care of the Galilei-invariance, it is
necessary to carefully check that the truncation of the $\eps$-series does
not introduce slight inaccuracies. Only after this has been done (for example
by re-summing the $\eps$-series) and checked by comparing with
non-perturbative data, meaningful quantitative statements on the 
scaling functions can be made. (iv) Throughout, it was implicitly assumed that
the order-parameter vanishes initially. Systematic studies on what
happens when this condition is relaxed are only now becoming available
\cite{Anni06,Cala06,Fedo06}. These extensions might be particularly important for chemical kinetics. (v) We did not include growth models here but
shall discuss them in section~4.
 
If $z=2$, it is also possible, using eq.~(\ref{gl:5:Crauschlos}), to derive
explicit predictions for the two-time correlation 
function \cite{Pico04,Henk04b}. These have been tested in some 
exactly solvable
models \cite{Pico04,Henk06a}, the $2D$ Ising model \cite{Henk04b} and the
$2D$ $q$-states Potts model with $q=2,3,8$ \cite{Lore06,Jank06}. Extensions
to $z=4$ have been studied very recently \cite{Roet06,Baum06e}, see
section~4.

This survey is organized as follows. In section~2 we review results on the
ageing behaviour of several critical models with non-equilibrium steady-states. 
The first two models are chosen because their steady-state 
phase-transitions are in the paradigmatic universality classes of
directed percolation (DP) and in the parity-conserving (PC) universality
classes. The numerical results on these models are supplemented by the
exactly solved bosonic variants of the contact and pair-contact processes. 
In this way it becomes clear that most aspects of the scaling description
of ageing in simple magnets does carry over to this more general situation. 
However, a central issue, namely the definition of an universal limit
fluctuation-dissipation ratio $X_{\infty}$ \cite{Godr00b} and 
which has received so much recent 
attention in the non-equilibrium critical dynamics of magnets, see \cite{Cris03,Cala05} for reviews, does not appear
to have an obvious analogue. Remarkably, the same kind of evidence in favour
of a non-trivial extension of dynamical scaling towards a larger dynamical
symmetry group of {\em local} scale-transformation previously found in 
magnets also appears in the models without detailed balance. 
In section~3, we review in more detail how the stochastic Langevin equations
underlying the bosonic contact and pair-contact processes can be shown
to actually possess a local scale-invariance. These evidences should form a
promising basis to look for more manifestation of local scale-invariance 
in systems with dynamical scaling which remain always very far from 
equilibrium. A different class of non-equilibrium models is studied in section~4, where kinetic growth as described by the Edwards-Wilkinson and the Mullins-Herring equations is studied. To account for those models, the
formulation of LSI must be generalized \cite{Henk02} to values $z\ne 2$ of the 
dynamical exponent. We consider explicitly the case $z=4$ and apply it to the Mullins-Herring model. We conclude in section~5. 
 
\section{Ageing with absorbing steady-states}

We now describe the ageing behaviour of system without an equilibrium
stationary state. We shall realize this system as models of interacting
classical particles, where the stochastic dynamics is such that the
detailed-balance condition no longer holds. For the simple models we shall
consider here it turns out that if the stationary state is not at a 
critical point, only a single stable stationary state remains to which
the system relaxes within a finite time and no ageing is possible.  
For this reason, we shall study the ageing behaviour at criticality. 

We remark that one might also go to non-equilibrium stationary states
by considering driven systems \cite{Schm95}. 
However, the dynamics of those is more
complicated than the systems at hand because of a further strong spatial
anisotropy and the description in terms of local scale-invariance would require
to generalize the local scale-transformations accordingly. That is beyond
the scope of this survey. Another very interesting class of ageing non-equilibrium
systems are zero-range process, see \cite{Evan05,Godr06} for recent reviews. Because they do not have a spatial structure
which would admit a Galilei-invariance, their dynamical scaling cannot be
extended to some form of local scale invariance and for this reason they are
not considered here, despite their intrinsic interest.

\begin{figure}
\centerline{
\epsfysize=50mm
\epsffile{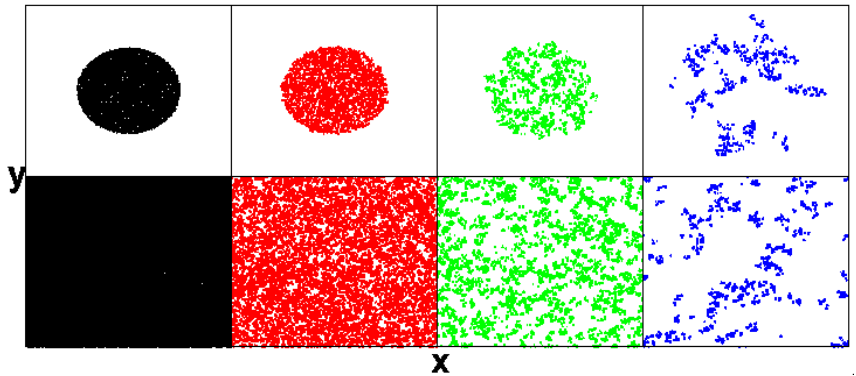}}
\caption[Evolution of clusters in the critical $2D$ contact process]{Microscopic evolution of clusters in the critical $2D$ contact 
process, on a lattice of size $1000 \times 1000$. 
The initial condition of the upper series is a full circle with radius $100$ 
placed in the center of the lattice, while in the lower series it is a full lattice. 
The times are $t = [2,20,200,2000]$ for the upper series
and $t =[20,200,2000,20000]$ below. After \cite{Rama04}. 
\label{abb:ndb:fig2}
}
\end{figure} 

\subsection{Contact process}

The contact process is a paradigmatic system for the study of
non-equilibrium phase-transitions, see e.g. \cite{Hinr00} for a 
review.\footnote{Recall that  bifurcations arising in many simple models
of mathematical biology \cite{Murr93} come from the mean-field treatment
of the phase-transition in the contact process.} The steady-state 
phase-transition of the contact process is in the same universality class
as one of the transitions of the celebrated Ziff-Gulari-Barshad model 
\cite{Ziff86}, which is meant to describe the catalytic reaction $2\mbox{\rm CO}+\mbox{\rm O$_2$} \longrightarrow 2\mbox{\rm CO$_2$}$. 
The model may be defined in terms
of a time-dependent discrete variable $n_i(t)\in\{0,1\}$, defined on each
site $i$ of a hypercubic lattice, which describe configurations of
particles and empty sites. The dynamics is defined as follows: for
each time-step, select randomly a site $i$ of the lattice. If $i$
is occupied (i.e. $n_i=1$), that particle vanishes with probability $p$. 
Otherwise, with probability $1-p$ a new particle is created on one of the
nearest neighbours of $i$, chosen at random and provided that chosen site
is still empty. Formally, this may be expressed through the reactions
$A\to\emptyset$ and $A\to2A$, with rates corresponding to $p$ and $1-p$,
respectively. In the steady-state, the model has a continuous phase-transition
at some critical value $p_c$. Numerically, $p_c=0.2326746(5)$ in $1D$ and
$p_c=0.37753(1)$ in $2D$. 

A first characteristic of the dynamics of the critical contact process 
can be seen by looking at the
temporal evolution of certain initial configurations, see figure~\ref{abb:ndb:fig2}. In contrast to magnetic systems, 
see figure~\ref{abb:amasc} for comparison, 
in the contact process there is no apparent growing length-scale at all and
the evolution proceeds via the slow dissolution of the particle clusters. 
Cluster dilution had first been demonstrated to occur in several variants of the
two-dimensional voter model \cite{Dorn01a} but also occurs in the early 
stages of surface ageing in simple magnets \cite{Plei04,Plei04a}.

In studies of the ageing behaviour, one goes beyond 
the average particle-density 
$N(t) := \langle n_i(t)\rangle\sim t^{-\delta}$ at criticality $p=p_c$. Define 
the two-time (connected) autocorrelator and autoresponse functions
\BEQ
C(t,s) := \langle n_i(t) n_i(s)\rangle - \langle n_i(t)\rangle \langle n_i(s)\rangle
\;\; , \;\; 
R(t,s) := \left.\frac{\delta\langle n_i(t)\rangle}{\delta h_i(s)}\right|_{h=0}
\EEQ
where $h_i(s)$ is the rate of the spontaneous creation process $\emptyset\to A$ at the site $i$ at time $s$. These may be calculated in a standard fashion 
either from simulations \cite{Rama04,Hinr06} or else from the 
transfer-matrix renormalization group \cite{Enss04}. Recent field-theoretical
calculations \cite{Baum06d} will also be described.

\begin{figure}
\centerline{
\epsfysize=45mm
\epsffile{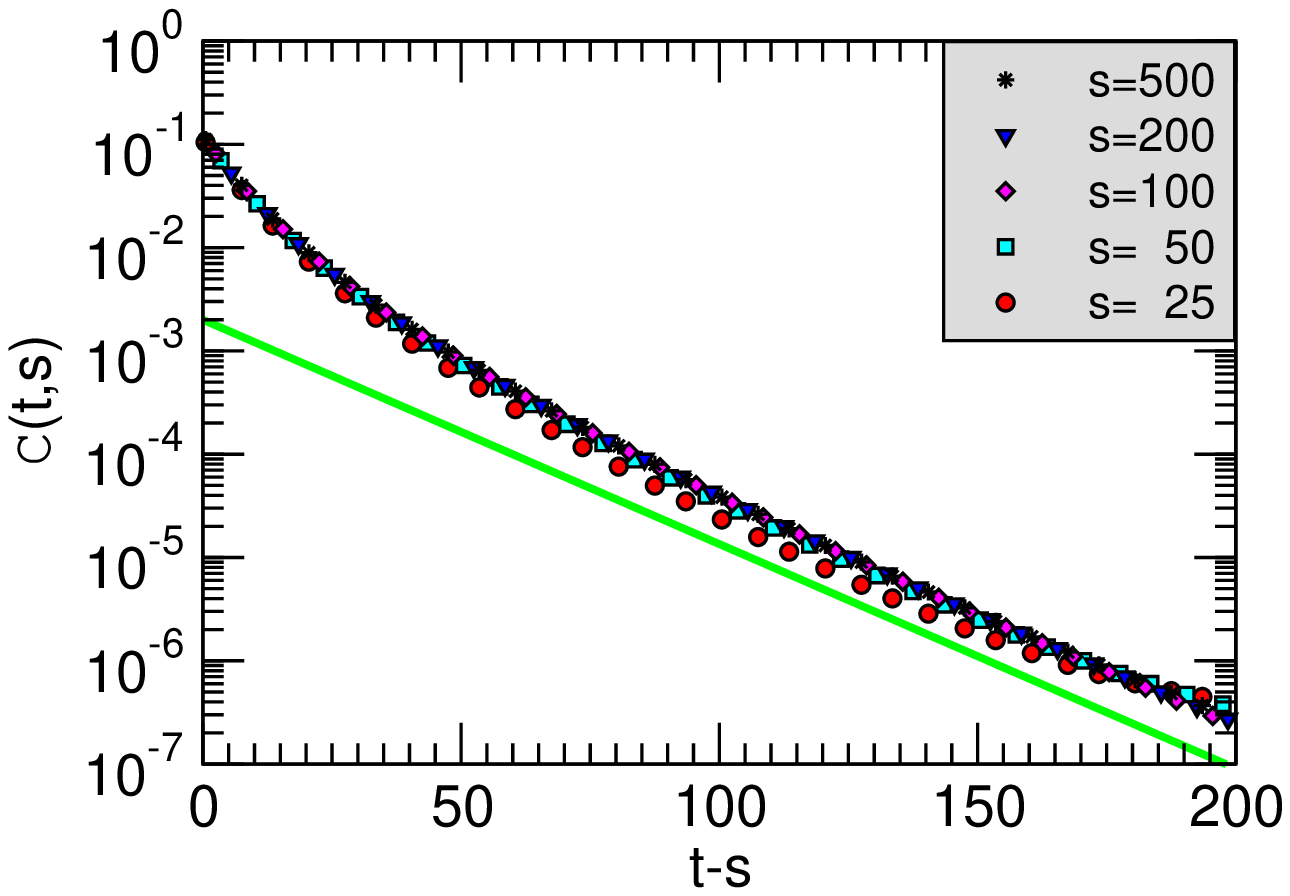}
\epsfysize=45mm
\epsffile{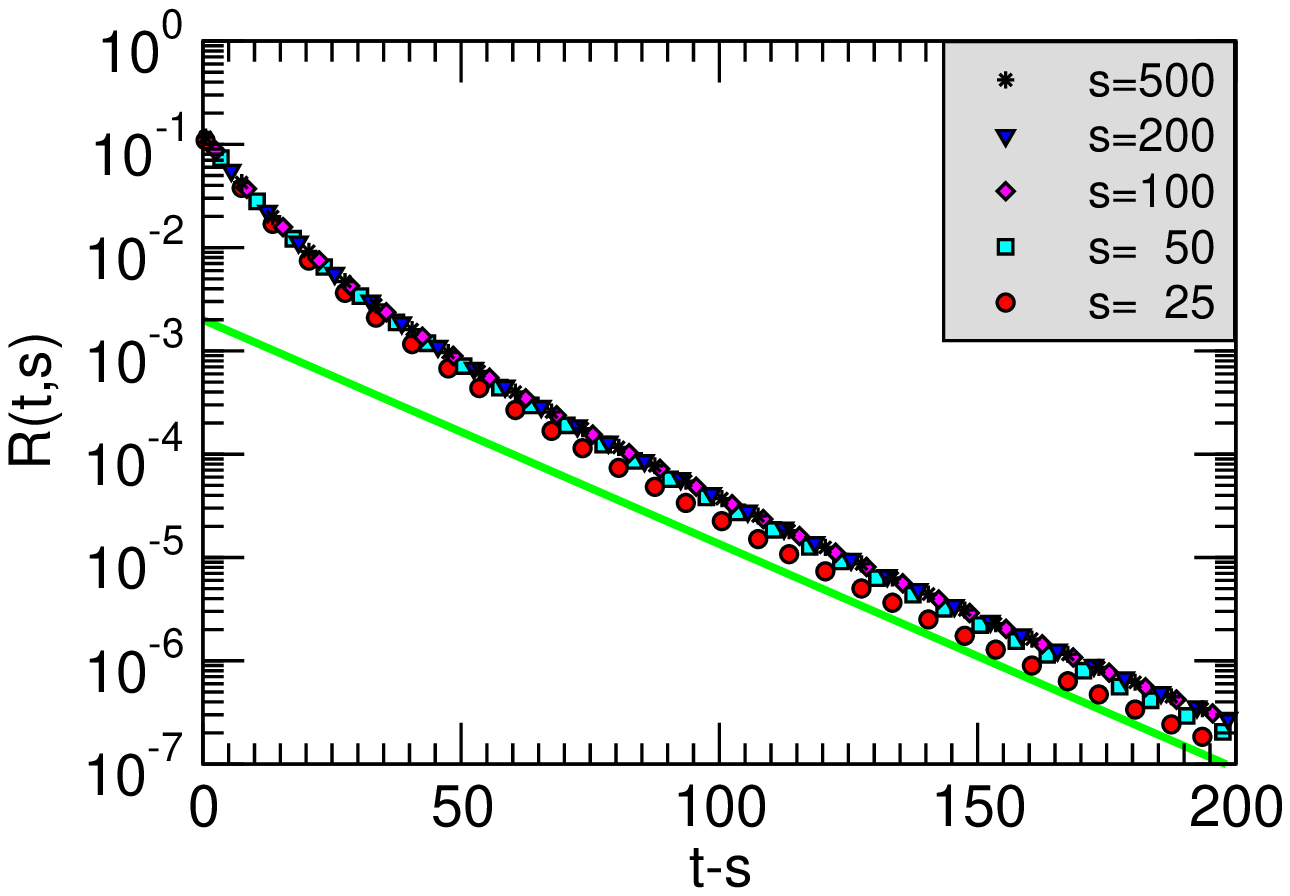}}
\caption[Two-time function in the active phase of the $1D$ contact process]{Connected autocorrelation function $C(t,s)$ and autoresponse function $R(t,s)$ of the $1D$ 
contact process in the active phase ($p=0.1$). The straight lines are proportional
to $\exp(-0.05 (t-s))$. After \cite{Enss04}. \label{abb:cpc:fig1314}}
\end{figure} 

\subsubsection{Active phase} 

In contrast with simple magnets, 
where there are
two distinct stable ground states in the low-temperature phase, in the active
phase of the contact process there is only a {\em single} stable steady state. 
Consequently, there is here {\em no} breaking of time-translation invariance and
we illustrate this in $1D$ in figure~\ref{abb:cpc:fig1314}. After a short 
transient, the data for both $C(t,s)$ and $R(t,s)$ collapse when plotted over
against $t-s$ which means that the contact process shows no ageing in its
active phase.

\subsubsection{Absorbing phase} 

{}From the comparison with the 
high-temperature phase of
simple magnets, one would also expect to find time-translation invariance in the
absorbing phase of the contact process. However, the correlation function shows
a subtlety the origin of which is best understood by considering the 
case $p=1$ first. 
If $p=1$, particles on different sites are uncorrelated and simply decay with a
fixed rate. For any fixed site $i$ and with two times $t>s$, it is clear that
$n_i(t) n_i(s) = n_i(t)$, since $n_i\in \{0,1\}$. Hence 
$\langle n_i(t) n_i(s)\rangle=\langle n_i(t)\rangle$ and $C(t,s)=N(t) (1-N(s))$. 
For sufficiently long times, $C(t,s)$ will then only depend on $t$. Indeed, this
behaviour survives in the entire absorbing phase \cite{Enss04}. 
On the other hand, the expected time-translation invariance for the autoresponse function is readily checked. 

\begin{figure}
\centerline{
\epsfysize=52mm
\epsffile{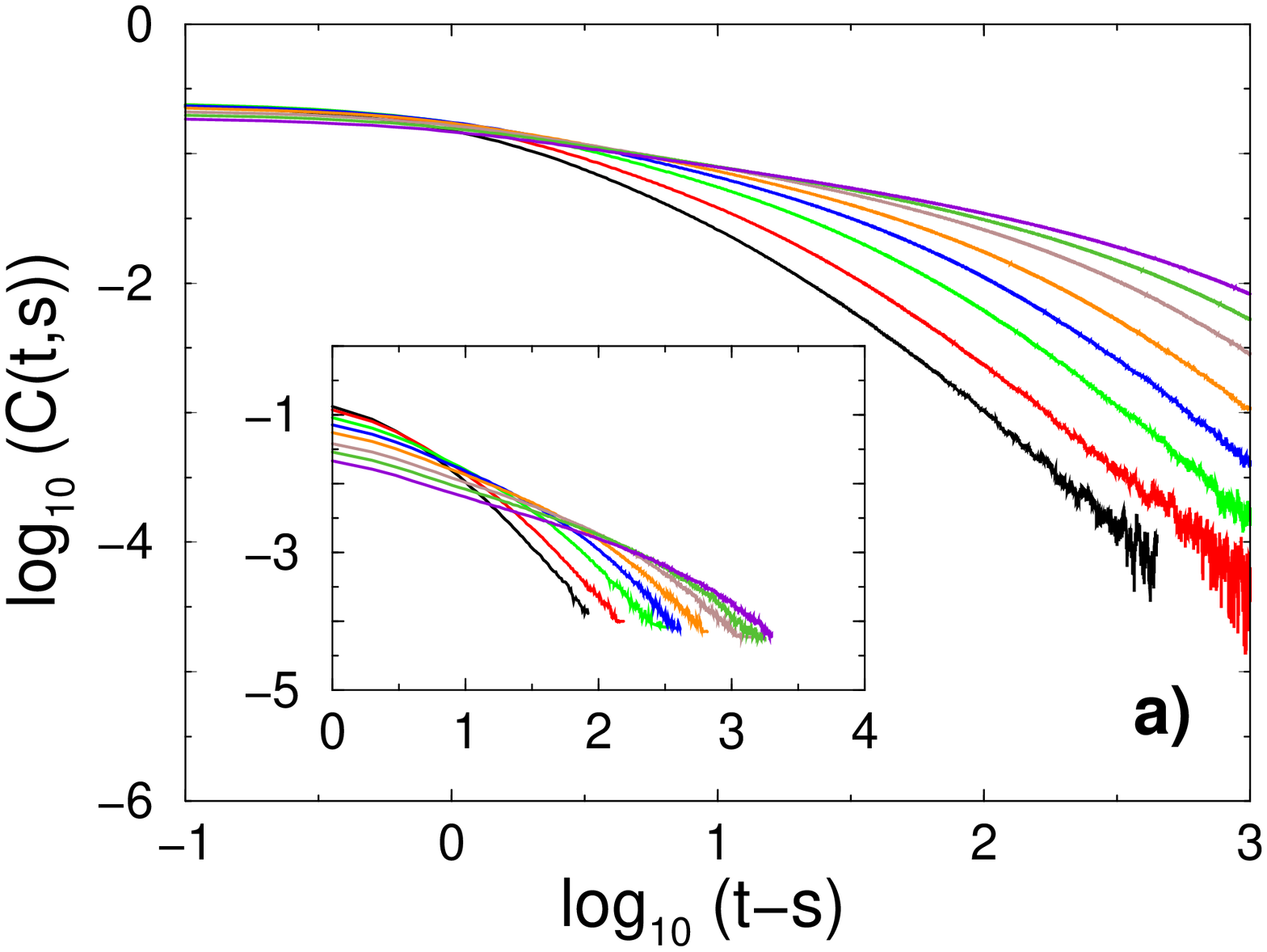}
\epsfysize=52mm
\epsffile{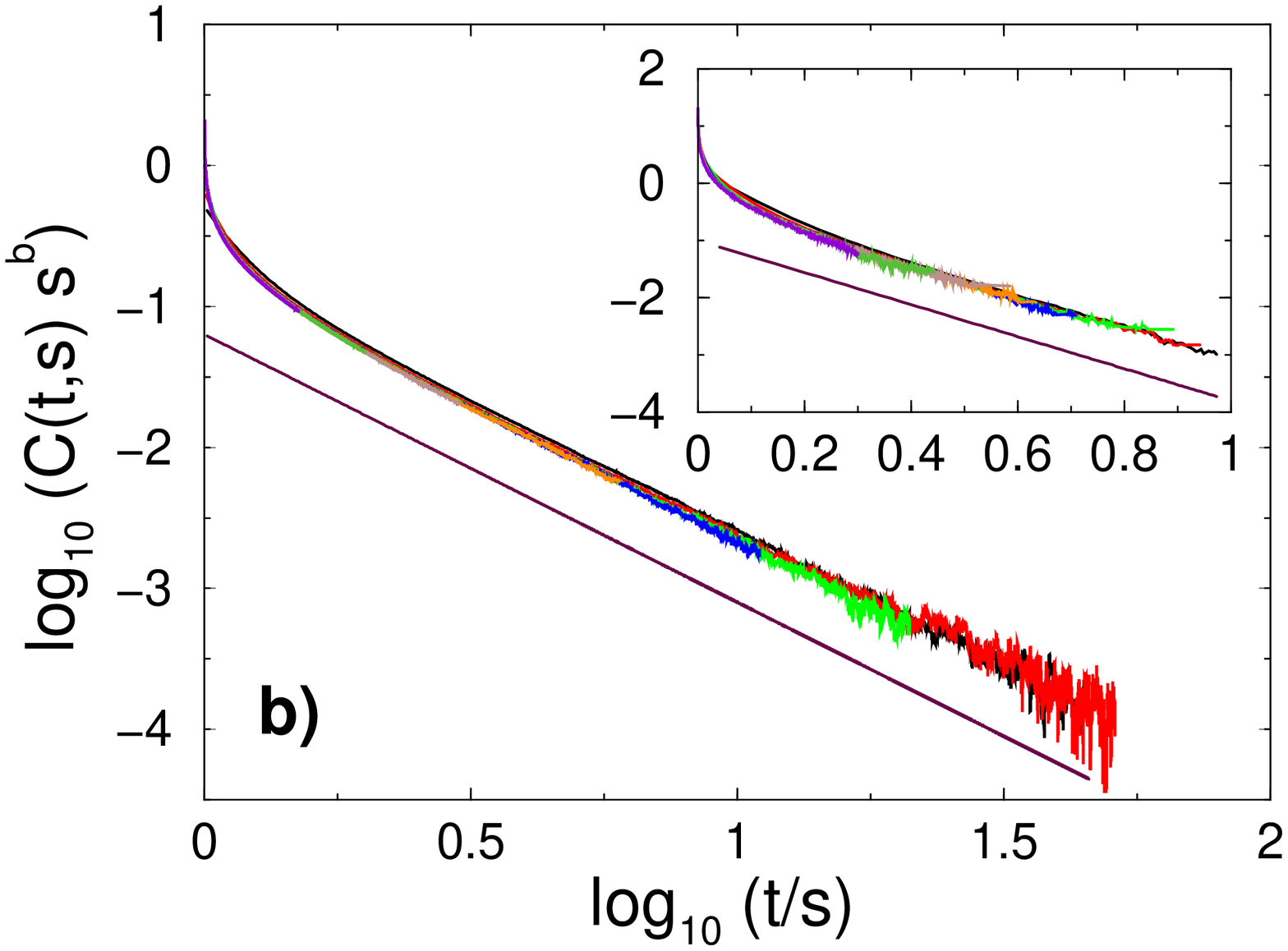}}
\caption[Connected autocorrelation function of the $1D$ and $2D$ critical
contact process]{Connected autocorrelation function of the critical
contact process in $1D$ (main plots) and $2D$ (insets). Panel (a) shows
the ageing of the autocorrelation function and panel (b) illustrates the scaling
behaviour. The straight lines correspond to the exponents 
$\lambda_C/z=1.9$ in $1D$ and $2.8$ in $2D$. After \cite{Rama04}.
\label{abb:cpc:fig3}}
\end{figure} 

\subsubsection{Critical point} 

\begin{figure}
\centerline{\epsfxsize=85mm\epsffile{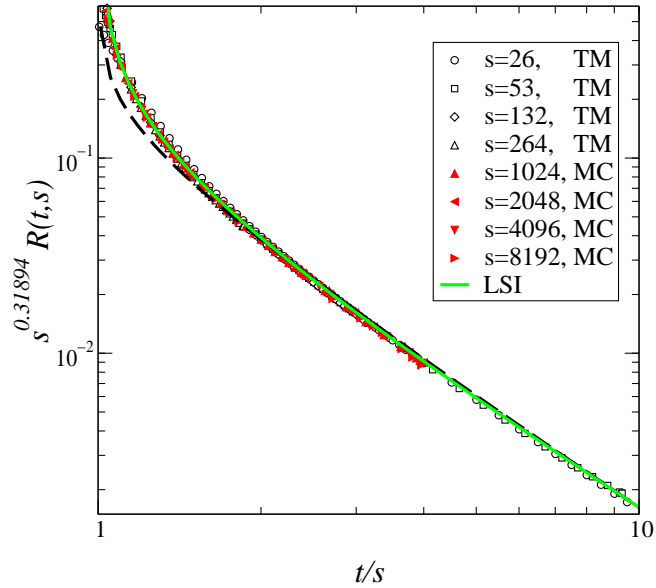}
}
\caption[Autoresponse function for the critical $1D$ contact 
process]{Autoresponse function for the critical $1D$
contact process for several waiting times $s$. The data labelled {\sc tm}
come from the transfer matrix renormalization group {\protect \cite{Enss04}} 
and {\sc mc} denotes Monte Carlo data. 
\label{abb:reposta:Abb2}
}
\end{figure}

For the critical contact process we show in 
figure~\ref{abb:cpc:fig3}a that ageing does occur, that is, the autocorrelation
and the autoresponse depend on {\em both} the observation time $t$ and the
waiting time $s$. Furthermore, when the same data are re-plotted over against 
$t/s$, a data collapse after rescaling can be achieved, 
see figure~\ref{abb:cpc:fig3}b. Lattices with a large initial particle
density $n\approx 0.8 -1$ were used \cite{Enss04,Rama04,Hinr06}. This is
different with respect to the magnetic systems of section~1, where the 
order-parameter had a vanishing initial value. 
By analogy with simple magnets, one defines the
ageing exponents $a,b$ and the autocorrelation and autoresponse 
exponents $\lambda_{C,R}$ from 
\begin{eqnarray}
C(t,s) &=& s^{-b} f_{C}(t/s) \;\; , \;\; 
f_{C}(y)\sim y^{-\lambda_{C}/z} \nonumber \\ 
R(t,s) &=& s^{-1-a} f_R(t/s) \;\; , \;\; f_R(y)\sim y^{-\lambda_R/z} 
\label{gl:cpc:ablCR}
\end{eqnarray}
where the asymptotic forms should hold for $y\to\infty$. 
Similarly, scaling can be observed for the autoresponse function as shown in
figure~\ref{abb:reposta:Abb2}. On the other hand, the unconnected correlator
behaves for large times simply as $\langle n_i(t)n_i(s)\rangle\sim(ts)^{-b/2}$. 

\begin{table}
\caption[Nonequilibrium exponents for fermionic and bosonic contact processes]
{Nonequilibrium exponents for the contact process ({\sc cp}), 
the non-equilibrium kinetic Ising model ({\sc nekim}), the 
bosonic contact process ({\sc bcp}) and the bosonic pair-contact process
({\sc bpcp}). Several kinetic growth models based on the Edwards-Wilkinson 
({\sc ew}) and Mullins-Herring ({\sc mh}) equations are also listed, see
section~4. 
 \label{tab:expoforaequi}}
\begin{center}
\begin{tabular}{|c|c|llll|cl|} \hline \hline\hline
model & $d$ & \multicolumn{1}{c}{$a$~} & \multicolumn{1}{c}{$b$~} 
    & $\lambda_{C}/z$ & $\lambda_R/z$ & method & Ref. \\ \hline\hline
{\sc cp} & 1 & $-0.68(5)$ & $0.32(5)$ & $1.85(10)$ & $1.85(10)$ &
    TMRG & \cite{Enss04} \\
 & & $-0.57(10)$ & $0.319$ & $1.9(1)$ & $1.9(1)$ &
    Monte Carlo & \cite{Rama04} \\
 & & $-0.681$    &         &          & $1.76(5)$ & 
    Monte Carlo & \cite{Hinr06} \\ \cline{2-8}
 & 2 & ~~~0.3(1) & $0.901(2)$ & $2.8(3)$ & $2.75(10)$ &
    Monte Carlo & \cite{Rama04} \\ \hline
{\sc nekim} & 1 & $-0.430(4)$ & $0.570(4)$ & 1.9(1) & $1.9(1)$ & Monte Carlo & \cite{Odor06} \\ \hline
{\sc bcp} & $\geq 1$ & $d/2-1$ & $d/2-1$ & $d/2$ & $d/2$ & exact & \cite{Baum05a} \\ \hline
{\sc bpcp} & $>2$ & $d/2-1$ & $d/2-1$ & $d/2$ & $d/2$ & exact, $\alpha<\alpha_C$ & 
\\ \cline{2-7}
  & $>2 ~\&~ <4$ & $d/2-1$ & 0 & $d/2$ & $d/2$ & exact, $\alpha=\alpha_C$ & \cite{Baum05a} 
\\ \cline{2-7}
  & $>4$ & $d/2-1$ & $d/2-2$ & $d/2$ & $d/2$ & exact, $\alpha=\alpha_C$ & 
\\ \hline\hline
{\sc ew2} & $\geq 1$ & $d/2-1$ & $d/2-1-\rho$ & $d/2-\rho$ & $d/2$ & exact & \cite{Roet06} \\ \hline
{\sc mh1} & $\geq 1$ & $d/4-1$ & $d/4-1$ & $d/4$ & $d/4$ & exact &\cite{Roet06}\\ \hline
{\sc mh2} & $\geq 1$ & $d/4-1$ & $d/4-1-\rho/2$ & $d/4-\rho/2$ & $d/4$ & exact &\cite{Roet06} \\ \hline
{\sc mh}c & $\geq 2$ & $(d-2)/4$ & $(d-2)/4$ & $(d+2)/4$ & $(d+2)/4$ & exact &
\cite{Baum06e} \\ 
\hline\hline\hline
\end{tabular}\end{center}
\end{table}

The results for the ageing exponents $a,b,\lambda_C,\lambda_R$ are collected in
table~\ref{tab:expoforaequi}. The agreement between the results of the TMRG and
Monte Carlo simulations serves as a useful contr\^ole on the fiability of the 
results. 
While the equality $\lambda_C=\lambda_R$ is fully analogous to what was seen 
in non-equilibrium critical dynamics, the exponents $a$ and $b$ are no longer 
equal but satisfy
\BEQ \label{gl:ab}
1+a = b = 2\delta
\EEQ
Some comments are in order.
\begin{enumerate}
\item If the critical contact process were a Markov process, these exponents 
might be calculated from the global persistence\index{persistence exponent} 
exponent $\theta_g$  through the 
scaling relation \cite{Maju96a,Hinr98,Alba01a} 
\begin{equation}
\frac{\lambda_C}{z} = \theta_g - \frac{2 (1-d)-\eta}{2 z} ,
\end{equation}
which would predict $\lambda_C/z = 1.98(2)$ in $1D$ and 
$\lambda_C/z = 3.5(5)$ in $2D$. Although this not too far form the
values reported in table~\ref{tab:expoforaequi}, the differences 
appear to be significant. If that conclusion is correct, it would point 
towards the existence of temporal 
long-range correlations and hence of an effective non-markovian dynamics
of the critical contact process.
\item The relation $1+a=b$ can be understood \cite{Baum06d} as follows to be a 
consequence of the rapidity-reversal symmetry of Reggeon field-theory (which 
is generally thought to be in the same universality class as the contact 
process). The field-theoretical action in the Janssen-de Dominics formulation
reads {\em at} the critical point
\BEQ
{\cal J}[\phi,\wit{\phi}] = \int\!\D t\D\vec{r}\: \left[
\wit{\phi}\left( \partial_t - D \Delta\right)\phi -
u\left(\wit{\phi}-\phi\right)\wit{\phi}\phi - h \wit{\phi} \right] 
\EEQ
where $\phi$ and $h$ are the coarse-grained particle-densities and 
creation rates for particles. For $h=0$ and if the ´time´ $t\in\mathbb{R}$
is unbounded, the action is invariant under the {\em rapidity reversal}, see
\cite{Taeu05,Taeu06}
\BEQ
\wit{\phi}(t,\vec{r}) \mapsto - \phi(-t,\vec{r}) \;\; , \;\;
{\phi}(t,\vec{r}) \mapsto - \wit{\phi}(-t,\vec{r})
\EEQ
In particular, it follows that the scaling dimensions
$x_{\phi}=x_{\wit{\phi}}=\beta/\nu$ must be equal. This remains true even if
rapidity-reversal is broken by initial conditions at time $t=0$. For 
a rapidity-reversal-invariant action $\cal J$, the connected correlator becomes
\cite{Baum06d}
\BEA
C(t,s;\vec{r}-\vec{r}') &=& \langle \phi(t,\vec{r})\phi(s,\vec{r}')\rangle 
- \langle \phi(t,\vec{r})\rangle \langle \phi(s,\vec{r}')\rangle 
\nonumber \\
&=& \langle \wit{\phi}(-t,\vec{r})\wit{\phi}(-s,\vec{r}')\rangle 
- \langle \wit{\phi}(-t,\vec{r})\rangle \langle \wit{\phi}(-s,\vec{r}')\rangle
\nonumber \\
&=& 0
\EEA
where the second line comes from rapidity-reversal symmetry and the 
last line follows from causality. Hence $C(t,s;\vec{r})=0$ in the steady-state
but for relaxations from an initial state $C(t,s;\vec{0})=\langle
\phi(t)\phi(s)\rangle_c$ and 
$R(t,s;\vec{0})=\langle \phi(t)\wit{\phi}(s)\rangle$
are {\em non}-vanishing and have the same scaling dimensions, which implies
(\ref{gl:ab}) and also $\lambda_C=\lambda_R$. 

While this explains the origin of eq.~(\ref{gl:ab}) for the contact
process, it is not yet understood why it also holds true in the
{\sc nekim} \cite{Odor06} (see below) where rapidity-reversal symmetry is not
known to be satisfied. On the other hand eq.~(\ref{gl:ab}) is not universally valid. In the critical {\em bosonic} contact process, one has $a=b$. The critical bosonic pair-contact process furnishes
further examples with $a\ne b$ but the relation between $a$ and $b$ is
distinct from eq.~(\ref{gl:ab}). 
\item The non-equality of the exponents $a$ and $b$ is only possible in
systems with non-equilibrium steady-states. Indeed, for equilibrium systems, 
one has time-translation invariance
and combining this with the scaling forms would give
\BD
C(t,s) \sim (t-s)^{-b} \;\; , \;\; R(t,s) \sim (t-s)^{-1-a}
\ED
The fluctuation-dissipation theorem would then give $a=b$. Hence the equality
$a=b$ is a necessary condition that a quasi-stationary state, which might 
be present for $t-s\ll 1$, is an equilibrium one. 
\item Is it possible to define a non-equilibrium temperature 
{}from the steady-states
of systems without detailed balance~? A recent attempt by Sastre, Dornic and 
Chat\'e \cite{Sast03} started from the observation that in simple magnets, the
fluctuation-dissipation ratio $X(t,s)\to 1$ as $t\to\infty$ and $t-s\to 0$. 
{}From this observation, they define a \emph{dynamical temperature} by 
\BEQ \label{gl:cpc:Tdyn}
\frac{1}{T_{\rm dyn}} := \lim_{t\to\infty}\left(\lim_{t-s\to 0}
\frac{R(t,s)}{\partial C(t,s)/\partial s}\right)
\EEQ 
By explicit calculation, they confirm that in the $2D$ critical voter model
(where indeed $a=b$) this limit exists, has a non-trivial value and is 
universal \cite{Sast03}. Still, their appealing idea has met with several 
criticisms. First, Mayer and Sollich \cite{Maye05a} construct 
in the coarsening $1D$ Glauber-Ising model a defect-pair observable 
such that the fluctuation-dissipation ratio $X(t,s)\ne 1$ in the short 
time-regime (in particular they show $\lim_{s\to\infty} X(s,s)=3/4$).
Second, $T_{\rm dyn}$ can only be finite if $a=b$ and the examples
listed in table~\ref{tab:expoforaequi} show that the
fluctuation-dissipation ratio itself is in general no longer defined. 
It appears that the proposal (\ref{gl:cpc:Tdyn}) 
relies too heavily on specific properties of the voter model. 
\item Rather than the fluctuation-dissipation ratio $X(t,s)$ 
as defined in eq.~(\ref{gl:rfd}) for systems with detailed balance, 
one may instead consider the ratio $\Xi(t,s)$, and its limit
$\Xi_{\infty}$, which are defined by \cite{Enss04}
\BEQ
\Xi(t,s) := \frac{R(t,s;\vec{0})}{C(t,s;\vec{0})} = 
\frac{f_R(t/s)}{f_C(t/s)} \;\; ; \;\;
\Xi_{\infty} := \lim_{s\to\infty}\left(\lim_{t\to\infty} \Xi(t,s)\right)
\EEQ 
which are well-defined because of (\ref{gl:ab}). The limit
$\Xi_{\infty}$, being the ratio of two quantities with the same classical
and scaling dimensions, is expected to be universal \cite{Baum06d}.
Baumann and Gambassi \cite{Baum06d} argue that a value $\Xi(t,s)^{-1}\ne 0$ 
is a measure for the breaking of the rapidity-reversal symmetry (in the same way
as $X\ne 1$ measures the distance from an equilibrium state) and show 
explicitly that only the zero-momentum modes contribute to a non-vanishing 
value of $\Xi(t,s)^{-1}$. In this respect, for systems with rapidity-reversal
symmetry, $\Xi$ appears to be the analogue of the fluctuation-dissipation 
ratio $R$. Of course, the applicability of their argument depends on the 
validity of the scaling relation $1+a=b$. 
For the value of $\Xi_{\infty}$, they find in $4-\eps$ dimensions
\BEQ \label{gl:Xiinfty}
\Xi_{\infty} = 2 \left[ 1 - \eps \left( \frac{119}{480}-\frac{\pi^2}{120}
\right)\right] + {\mbox\rm O}(\eps^2)
\EEQ
which in $1D$ is in semiquantitative agreement with the estimate 
$\Xi_{\infty}=1.15(5)$ \cite{Enss04}.  
\end{enumerate} 

Before one can discuss the form of the scaling function $f_R(y)$, it is
necessary to study the r\^ole of the initial conditions. Indeed, all
existing simulations on the ageing in that model start from a lattice with a 
non-vanishing particle density and hence in contrast to 
simulations in magnetic system, where
the initial order-parameter was set to zero. For the time-dependent
order-parameter one expects the scaling 
form \cite{Wijl98,Jans89,Jans92,Baum06d}
\BEQ
\langle \phi(t)\rangle = \phi_0 t^{\theta} 
f\left(\phi_0 t^{\theta +\beta/(\nu z)}\right)
\EEQ
where $\phi_0=\langle \phi(0)\rangle$ is the initial value of the
order-parameter (which for the contact process is the particle density) and
$\theta$ is the slip exponent. Hence there is characteristic time scale
$\tau_{*}\sim \phi_0^{-1/(\theta+\beta/(\nu z))}$ where a change of scaling
behaviour takes place. For any non-vanishing value of the dimensionful 
variable $\phi_0$ the long-time
scaling behaviour is effectively described by the $\tau_{*}\to 0$ limit.
For a vanishing initial order-parameter the exponents
$\lambda_{C,R}$ are independent of the equilibrium exponents \cite{Jans89}.  
For $\phi_0\ne 0$, Baumann and Gambassi show
from an analysis of the scaling behaviour of both responses and correlators 
that \cite{Baum06d}
\BEQ \label{gl:BaGa}
{\lambda_C} = {\lambda_R} = {d+z} + \frac{\beta}{\nu}
\EEQ  
The available exponent estimates in $1D$ and $2D$ from simulational studies
\cite{Enss04,Rama04,Hinr06,Henk06a} agree quite well with this prediction. 
We point out that the available results for the limit ratio $X_{\infty}$ 
\cite{Enss04,Baum06d}, see eq.~(\ref{gl:Xiinfty}), are found for a finite
initial particle-density. 

On the other hand, in the theory of LSI as reviewed in section~1 
the implicit assumption
$\phi_0=0$ was made. Since $\phi_0$ couples to a relevant scaling variable, 
it is not obvious why a theory formulated in the limit
$\phi_0 t^{\theta+\beta/(\nu z)}\to 0$ should be valid in the opposite
limit $\phi_0 t^{\theta+\beta/(\nu z)}\to \infty$ where {\em all} existing
studies on the ageing in the contact process have been performed. It is
therefore remarkable that LSI with $\phi_0=0$ still captures well the behaviour
of the $\phi_0\ne 0$ data. This is shown in figure~\ref{abb:reposta:Abb2},
where for almost all values of $t/s$ an almost perfect agreement with the 
LSI prediction (\ref{gl:Rf}), derived for $\phi_0=0$, is found.\footnote{In the
limit $\phi_0\to 0$, the results for $R(t,s)$ appear to be consistent with
LSI \cite{Baum06d}.} 
Hinrichsen \cite{Hinr06} has argued
that a more ambitious test may be performed by plotting
$r(y) := f_R(y) [y^{\lambda_R/z-1-a'} (y-1)^{1+a'}]$ over against $t/s-1$. 
If LSI with $\phi_0=0$ holds, one expects that $r(y)={\rm const.}$  
In this way, he found that for although the measured function 
$f_R(y)$ agrees nicely with eq.~(\ref{gl:Rf}) if $t/s$ is large enough, 
for values $t/s\lesssim 1.1$, the function $f_R(y)$ remains well-defined but 
changes to a different behaviour which is no longer described by
eq.~(\ref{gl:Rf}). The high
quality of his data makes it clear that this change of behaviour in the
scaling function cannot be explained away by invoking corrections to scaling. 
Further unpublished calculations \cite{Enss06} for extremely large values of
$s$ confirm these conclusions. In addition, the same conclusion has been 
reached from a detailed field-theoretic study of the two-time response in 
momentum space \cite{Baum06d}. A possible explanation of the form of $f_R(y)$ 
in terms of LSI will require the extension of the theory to include a 
non-vanishing value of $\phi_0$.

\subsection{Non-equilibrium kinetic Ising model}
 
We now review results, obtained by \'Odor \cite{Odor06}, on the ageing 
behaviour in a non-equilibrium kinetic Ising model ({\sc nekim}) 
where the parity of the 
total particle-number is conserved. The model was introduced by
Menyh\'ard \cite{Meny94} and combines spin-flips as in zero-temperature 
Glauber dynamics with
spin-exchanges as in Kawasaki dynamics. The model is formulated either 
in terms of Ising spins ($\uparrow$, $\downarrow$) or else in terms of a 
particle-reaction model of the kinks with occupied or empty sites 
($\bullet$, $\circ$). 
First, the Glauber-like part of the dynamics contains a diffusive motion
\begin{displaymath}
\uparrow\downarrow\downarrow\stackrel{}{\rightleftharpoons}
\uparrow\uparrow\downarrow \ \
\mbox{\rm ~or equivalently~} \ \ \bullet\circ \rightleftharpoons \circ\bullet
\;\; ; \;\; \mbox{\rm with rate $D$} 
\end{displaymath}
and the pair-annihilation of nearest neighbours
\begin{displaymath}
\uparrow\downarrow\uparrow \stackrel{}{\rightarrow}
\uparrow\uparrow\uparrow
 \ \ \mbox{\rm ~or equivalently~} \ \ \bullet\bullet \rightarrow \circ\circ
\;\; ; \;\; \mbox{\rm with rate $2\alpha$} 
\end{displaymath}
The Kawasaki-like part of the dynamics is described by
\begin{displaymath}
\uparrow\uparrow\downarrow\downarrow\stackrel{}{\rightleftharpoons}
\uparrow\downarrow\uparrow\downarrow \ \
\mbox{\rm ~or equivalently~} \ \ \circ\bullet\circ \rightleftharpoons 
\bullet\bullet\bullet
\;\; ; \;\; \mbox{\rm with rate $k$} 
\end{displaymath}
In full, this is a model describing branching and annihilating random walks
with an even number of offspring. By increasing $k$, one finds a
second-order phase-transition \cite{Meny94} where the kinks go from an 
absorbing to an active state. This phase-transition is in the so-called
parity-conserving (PC) universality class \cite{Gras84,Card96,Cane05} which 
is different from the one of the contact process. 

Using the parameterization $k=1-2\Gamma$, $D=\Gamma(1-\bar{\delta})/2$ and $
2\alpha=\Gamma(1+\bar{\delta})$, the critical point is located at
$\Gamma=0.35$, $k=0.3$ and $\bar{\delta}=-0.3928$ \cite{Odor06}. Measuring the
kink-density through an efficient cluster algorithm, and starting from a
fully ordered kink state with spins being alternatingly $\uparrow$ and
$\downarrow$, he finds  a nice
power-law scaling $\langle n_i(t)\rangle \sim t^{-0.285(2)}$. 

Next, measuring the unconnected kink-kink two-time correlation 
function, \'Odor's data are fully compatible with the scaling behaviour 
$\langle n_i(t)n_i(t')\rangle \sim t'^{-0.57} (t/t')^{-0.285}$ 
for $t/t'\to\infty$ and allows to determine the 
ageing exponent $b$, see table~\ref{tab:expoforaequi}. 
This result is also consistent with earlier results on the spin-spin 
autocorrelator in the same model, see \cite{Odor06} for details. The connected 
autocorrelator was also calculated, with the result
$\lambda_C/z=\lambda_R/z=1.9(1)$. This is in agreement with the scaling
relation (\ref{gl:BaGa}).  

\begin{figure}
\centerline{\epsfxsize=95mm\epsffile{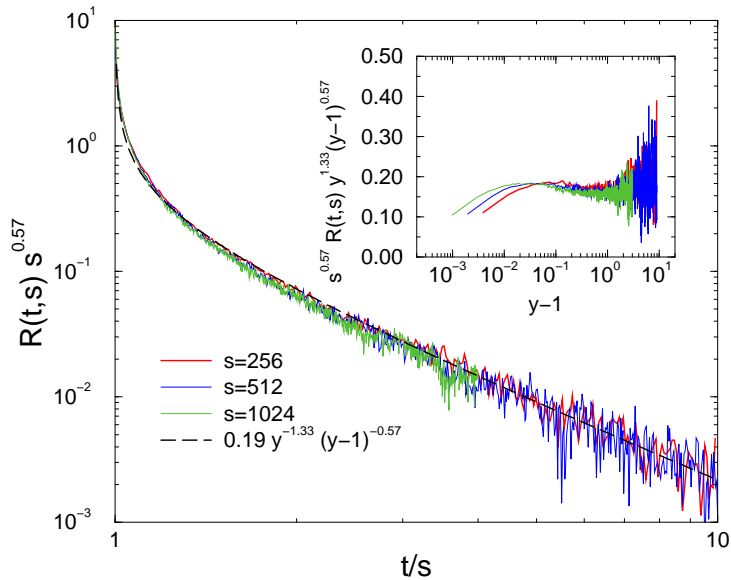}
}
\caption[Autoresponse function for the critical $1D$ NEKIM]{Scaling of the 
autoresponse function for the critical $1D$ {\sc nekim} \cite{Odor06}, 
for several waiting times $s$. The inset shows a rescaled response
function, with $s=[256,512,1024]$ from right to left.
\label{abb:Geza}
}
\end{figure}

Finally, the spin-autoresponse $R(t,s)$ with respect to a magnetic field 
coupling to a spin was calculated \cite{Odor06}, 
by adapting methods used previously
by Hinrichsen \cite{Hinr06} for the contact process. We recall that for the
fully ordered kink initial state, the initial spin magnetization vanishes. 
In figure~\ref{abb:Geza} it is shown that a clear data collapse is 
found when $1+a=b$, as in the contact process. 

Comparing the form of the scaling function $f_R(y)$ with the prediction of
LSI, \'Odor finds a perfect agreement as long as his data remain in the scaling 
limit. In particular, he carefully considered the limit $y\to 1$ (see
inset in figure~\ref{abb:Geza}). Within the numerical accuracy, 
the data for the rescaled 
response scaling function $r(y)$ remain essentially constant for all values
of $t/s$, as long the model is in the scaling regime. Clear deviations from a
horizontal line only occur once finite-time corrections have broken dynamical 
scaling and by increasing
$s$, the regime where (i) scaling holds and (ii) $r(y)$ 
is constant progressively
extends to ever smaller values of $y=t/s$. This finding
is different from the $1D$ contact process \cite{Hinr06,Baum06d} 
(see above). It is conceivable that the agreement with LSI in the
{\sc nekim} comes from the fact that the initial configuration used has
a vanishing magnetization. 

\subsection{Bosonic contact and pair-contact processes}

Two exactly solvable models allow to confirm the conclusions drawn
{}from the above numerical studies. 

The bosonic contact process was introduced in order 
to describe clustering phenomena in biological systems \cite{Houc02} whereas
the bosonic pair-contact process was originally conceived \cite{Paes04} 
as a solvable variant of the usual (fermionic) pair-contact process. These 
models are defined as follows. Consider a set of particles of a single species 
$A$ which move on the sites of a $d$-dimensional hypercubic lattice. On any site
one may have an arbitrary (non-negative) number of particles and it is this 
property which makes
up the difference with the usual contact and pair-contact processes considered
before where on each site at most one particle is allowed.  
Single particles may hop to a nearest-neighbour site with unit rate and in 
addition, the following single-site creation and annihilation processes 
are admitted 
\BEQ \label{gl:4:bcp:rates}
m A \stackrel{\mu}{\longrightarrow} (m+1) A \;\; , \;\; 
p A \stackrel{\lambda}{\longrightarrow} (p-\ell) A \;\; ; \;\;
\mbox{\rm with rates $\mu$ and $\lambda$} 
\EEQ
where $\ell$ is a positive integer such that $|\ell| \leq p$. 
We are interested in the following special cases:
\begin{enumerate}
\item {\em critical bosonic contact process:}\index{bosonic contact process} 
$p=m=1$. Here only $\ell = 1$ is possible. Furthermore
the creation and annihilation rates are set equal $\mu =\lambda$.
\item {\em critical bosonic pair-contact 
process:}\index{bosonic pair-contact process} $p=m=2$. We fix
$\ell=2$, set $2 \lambda = \mu$ and define the control parameter
\footnote{If instead we would treat a coagulation process $2A\to A$, 
where $\ell=1$, the results presented in the text are recovered by setting
$\lambda = \mu$ and $\alpha = {\mu}/{D}$.}
\BEQ
\alpha := \frac{3 \mu}{2 D}
\EEQ
\end{enumerate}

The dynamics of these models is conveniently described in terms creation 
operators $a^{\dag}(t,\vec{r})$ of a particle at time $t$ and location 
$\vec{r}$ and the
corresponding annihilation operator $a(t,\vec{r})$, see \cite{Doi76,Schu00}. 
The equation of motion for the space-time-dependent
particle-density $\rho(t,\vec{x}) := \langle
a^{\dag}(t,\vec{x}) a(t,\vec{x}) \rangle = \langle
a(t,\vec{x}) \rangle $ reads, after a rescaling $t\mapsto t/(2D)$ 
\cite{Houc02,Paes04} 
\BEQ
\hspace{-1truecm} \frac{\partial}{\partial t} 
\left\langle a(t,\vec{x})\right\rangle 
= \frac{1}{2} \Delta_{\vec{x}} \left\langle a(t,\vec{x})\right\rangle 
-\frac{\lambda\ell}{2D} \left\langle a(t,\vec{x})^p\right\rangle
+\frac{\mu}{2D} \left\langle a(t,\vec{x})^m\right\rangle + h(t,\vec{x})
\EEQ
where we have used the short-hand
\BEQ
\Delta_{\vec{x}} f(t,\vec{x}) := \sum_{r=1}^{d}  
\Bigl( f(t,\vec{x}-\vec{e}_r) + f(t,\vec{x}+\vec{e}_r)
-2f(t,\vec{x}) \Bigr)
\EEQ
Similarly, the equal-time correlation functions satisfy the equations of motion 
\BEA
\hspace{-1truecm} \frac{\partial}{\partial t} \expect{a(\vec{x})a(\vec{y})} 
&\hspace{-0.1truecm}=&
\hspace{-0.0truecm}\frac{1}{2}\sum_{k=1}^{d} \Bigl[ 
                   \expect{ a(\vec{x})a(\vec{y}-\vec{k}) } + 
                   \expect{ a(\vec{x})a(\vec{y}+\vec{k}) }
\nonumber 
\label{gl:4:bcp:movimento1}\\
& &  \hspace{-0.2truecm}+ \expect{ a(\vec{x}-\vec{k})a(\vec{y}) } +
             \expect{ a(\vec{x}+\vec{k})a(\vec{y}) } -
           4 \expect{ a(\vec{x})a(\vec{y})} \Bigr]
\\
\hspace{-1truecm}\frac{\partial}{\partial t} \expect{\left(a(\vec{x})\right)^2} 
&\hspace{-0.1truecm}=& 
\hspace{-0.0truecm} \sum_{k=1}^{d} 
                    \Bigl[ \expect{ a(\vec{x})a(\vec{x}-\vec{k}) } +
                            \expect{ a(\vec{x})a(\vec{x}+\vec{k}) } -
                              2\expect{ a(\vec{x})^2} \Bigr] 
\nonumber \\
& & \hspace{-0.0truecm}+\frac{\mu (1+\ell)}{2D} \expect{a(\vec{x})^m} 
\label{gl:4:bcp:movimento2}
\EEA
where in eq.~(\ref{gl:4:bcp:movimento1}) 
$\vec{x}\ne\vec{y}$ is understood. 
Since $\langle n(\vec{x})^2\rangle = \langle a(\vec{x})^2\rangle+ 
\langle a(\vec{x})\rangle$, the main equal-time quantity of interest, namely
the variance $\sigma^2 :=\langle n(\vec{x})^2\rangle
-\langle n(\vec{x})\rangle^2$
can be found. 

\begin{figure}[htb] 
  \vspace{0.5cm}
  \centerline{\epsfxsize=5.0in\epsfclipon\epsfbox
  {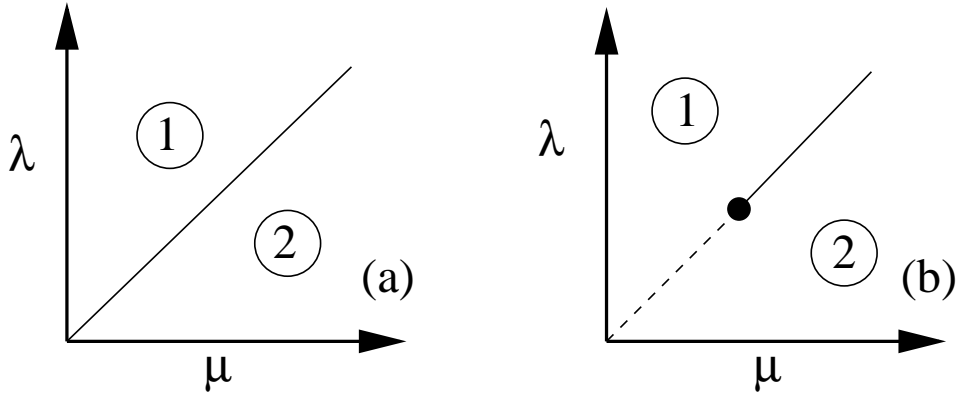}
  }
  \caption[Phase diagram of the bosonic contact and pair-contact 
  processes]{Schematic  
  phase-diagrams for $D\ne 0$ of (a) 
  the bosonic contact process and the bosonic pair-contact process in $d\leq 2$ 
  dimensions and (b)
  the bosonic pair-contact process in $d>2$ dimensions. The absorbing region 1, 
  where $\lim_{t \rightarrow \infty} \rho(\vec{x},t)= 0$, is separated
  by the critical line eq.~(\ref{gl:4:bcp:Krit}) from the active region 2, 
  where $\rho(\vec{x},t)\to\infty$ as $t\to\infty$. Clustering along
  the critical line is indicated in (a) and (b) by full lines, 
  but in the bosonic pair-contact process with $d>2$ the steady-state may also 
  be homogeneous (broken line in (b)). These two regimes are separated by a
  multicritical point. 
  \label{abb:bcp:Abb0}}
\end{figure}

The equations of motion (\ref{gl:4:bcp:movimento1},\ref{gl:4:bcp:movimento2})
are already written for the critical line given by \cite{Paes04} 
\BEQ \label{gl:4:bcp:Krit}
\ell \lambda = \mu.
\EEQ
where they naturally close. For the bosonic contact process $p=m=1$ there is an extension to arbitrary values of $\lambda,\mu$ which still closes. In both 
models, the spatial average of the 
local particle-density $\rho(\vec{x},t) := \langle a(\vec{x},t) \rangle$. 
remains constant in time
\BEQ
\int\!\D\vec{x}\, \rho(\vec{x},t) = 
\int\!\D\vec{x}\, \langle a(\vec{x},t) \rangle = \rho_0 
\EEQ
where $\rho_0$ is the initial mean particle-density. 
Furthermore, the critical line (\ref{gl:4:bcp:Krit}) 
separates an active phase with a 
formally infinite particle-density
in the steady-state from an absorbing phase where 
the steady-state particle-density vanishes. The phase diagrams are 
sketched in figure~\ref{abb:bcp:Abb0}. 
 
The physical nature of this transition becomes apparent when 
equal-time correlations are studied \cite{Houc02,Paes04}. 
For example, for the bosonic contact process at criticality 
one has \cite{Houc02}
\begin{equation}
\expect{a(t,\vec{x})^2}= \left\{ \begin{array}{ll}
      c_1 \,\, t^{-d/2+1}       & \mbox{\rm ~~;~ if $d<2$} \\
      c_2 \,\, \ln t            & \mbox{\rm ~~;~ if $d=2$} \\
      c_3 + c_4 \,\, t^{-d/2+1} & \mbox{\rm ~~;~ if $d>2$} 
\end{array} \right.
\end{equation}
where $t\gg 1$ and $c_0, \ldots, c_4$ are known 
positive constants. For $d\leq 2$, the
{\em fluctuations} in the mean particle-density increases with time, although
the mean particle-density itself remains constant. Physically, this means that
the particle-number on relatively few sites will increase 
while many other sites
will become empty. Only for $d>2$ fluctuations will eventually die out. 
For the bosonic contact process, this critical behaviour is the same along the
entire critical line. 

For the bosonic pair-contact process, that is different. Rather, there exists
a critical value $\alpha_C$ of the control parameter, given by 
\BEQ
\frac{1}{\alpha_C} = 2 \int_0^{\infty} \!\D u\, \left( e^{-4u} I_0(4u)\right)^d
\label{gl:def_alphaC}
\EEQ
and where $I_0(u)$ is a modified Bessel function. Specific values are
$\alpha_C(3) \approx 3.99$ and $\alpha_C(4) \approx 6.45$ and
$\lim_{d\,\searrow\, 2}\alpha_C(d)=0$. 
Then the variance behaves as \cite{Paes04}
\BEQ
\expect{a(t,\vec{x})^2}= \left\{ \begin{array}{ll}
f_0              & \mbox{\rm ~~;~ if $\alpha<\alpha_C$} \\
f_1\: t^{d/2-1}  & \mbox{\rm ~~;~ if $\alpha=\alpha_C$ and $2<d<4$} \\
f_2\: t          & \mbox{\rm ~~;~ if $\alpha=\alpha_C$ and $d>4$} \\
f_3\: e^{t/\tau} & \mbox{\rm ~~;~ if $\alpha>\alpha_C$ (or $d<2$)} 
\end{array} \right.
\EEQ
where the $f_0,\ldots,f_3$ and $\tau$ are known positive constants. 
This means that at the multicritical point at $\alpha=\alpha_C$ there occurs a 
{clustering transition} such that for $\alpha<\alpha_C$ the systems evolves
towards a more or less homogeneous state while for 
$\alpha\geq \alpha_C$ particles
accumulate on very few lattice sites while the other ones remain empty. 
In contrast with the bosonic contact process, clustering occurs in some region
of the parameter space for all values of $d$. 

We are interested in studying the impact of this clustering transition on the 
two-time correlations and linear responses. 
In order to obtain the equations of motion of the two-time correlator, the 
time-ordering of the operators $a(t,\vec{x})$ must be taken in account. 
This leads to the following equations of 
motion for the two-time correlator, after rescaling the 
times $t\mapsto t/(2D)$, $s\mapsto s/(2D)$, and for $t>s$,
\cite{Glau63,Baum05a}
\BEA
\label{gl:4:bcp:movimento3} 
& & ~~~~ \frac{\partial}{\partial t} 
\left\langle a(t,\vec{x}) a(s,\vec{y})\right\rangle 
\\
&=& \frac{1}{2} \Delta_{\vec{x}} \left\langle a(t,\vec{x}) 
a(s,\vec{y})\right\rangle
-\frac{\lambda\ell}{2D} \left\langle a(t,\vec{x})^p a(s,\vec{y})\right\rangle  
+\frac{\mu}{2D} \left\langle a(t,\vec{x})^m a(s,\vec{y})\right\rangle
\nonumber 
\EEA 
We are interested in the two-time connected correlation function\footnote{It can
be shown that in the scaling regime $C(t,s;\vec{r})\simeq \langle n(t,\vec{r}_0)
n(s,\vec{r}+\vec{r}_0)\rangle$ describes the two-time density-density
correlation function \cite{Baum06c}.}
\BEQ
\label{gl:4:bcp:C2}
C(t,s;\vec{r}) := \langle a(t,\vec{x}) a(s,\vec{x}+\vec{r})\rangle - \rho_0^2
\EEQ
and take an uncorrelated initial state, hence $C(0,0;\vec{r})=0$. The linear 
two-time response function is found by adding a particle-creation term 
$\sum_{\vec{x}} h(\vec{x},t)\left(a^\dag(\vec{x},t)-1\right)$ 
to the quantum hamiltonian $H$ and taking the functional derivative
\BEQ \label{gl:4:bcp:R2}
R(t,s;\vec{r}) := 
\left.\frac{\delta \langle a(t,\vec{r}+\vec{x})\rangle}{\delta h(s,\vec{x})}
\right|_{h=0}
\EEQ
and for which the usual scaling behaviour (\ref{gl:cpc:ablCR}) is anticipated. 

The solution of the equations (\ref{gl:4:bcp:movimento3}) for the two-time
quantities is now straightforward, if just a little tedious. It can be shown 
\cite{Baum05a} that the anticipated scaling behaviour (\ref{gl:cpc:ablCR})
exists along the critical line, but for the {\sc bpcp} the
further condition $\alpha\leq\alpha_C$ is required. For these
cases, the exponents are listed in table~\ref{tab:expoforaequi}. In
particular, we see that at the multicritical point $\alpha=\alpha_C$, the
exponents $a$ and $b$ are different and furthermore, do {\em not} satisfy the
relation $1+a=b$ found for the critical contact process and the {\sc nekim}. 
This means that there is no straightforward way to define an analogue of a
limit fluctuation-dissipation ratio for particle-reaction models 
without detailed balance. 
Furthermore, the explicit form of the scaling functions can also be
found and are listed in table~\ref{tb:results_fun}. 
While the form of the autoresponse function $f_R(y)=(y-1)^{-d/2}$ 
is remarkably simple, the results
for the autocorrelation function can be rendered as an integral
\BEQ
\label{gl:4:bcp:formaintegralC}
f_C(y) = \mathcal{C}_0 \int_0^1 \!\D\theta\, 
\theta^{a-b} (y+1-2 \theta)^{-{d}/{2}}
\EEQ
where the exponents $a,b$ are taken from table~\ref{tab:expoforaequi}. 

In section~3, we shall show how these results for $f_R(y)$ and $f_C(y)$ in the 
{\sc bcp} and the {\sc bpcp} can be understood using local scale-invariance. 
In section~4, we shall define the {\sc ew} and {\sc mh} models and, after having briefly reviewed LSI for $z=4$, then perform a similar analysis. 

\begin{table} 
  \[
  \hspace*{-2.5cm}
  \begin{array}{||c|c|c||c|c||}  \hline \hline
    \multicolumn{3}{||c||}{} &  f_R(y) & f_C(y) \\
    \hline
    \hline
    \multicolumn{3}{||c||}{\mbox{contact process {\sc bcp}}} &
    (y-1)^{-\frac{d}{2}}  & (y-1)^{-\frac{d}{2}+1} -
    (y+1)^{-\frac{d}{2}+1} \\ \hline \hline
    \mbox{pair} & \alpha < \alpha_C & d > 2 & (y-1)^{-\frac{d}{2}} &
    (y-1)^{-\frac{d}{2}+1} - (y+1)^{-\frac{d}{2}+1}  \\ \cline{2-5}
    \mbox{contact} &  & 2 < d < 4 & (y-1)^{-\frac{d}{2}} & 
    (y+1)^{-\frac{d}{2}} {_2F_1}
    \left(\frac{d}{2},\frac{d}{2};\frac{d}{2}+1;\frac{2}{y+1}\right)\\
    \cline{3-5}
    \mbox{process} & \raisebox{1.6ex}[-1.6ex]{$\alpha =
    \alpha_C$} & d > 4 & (y-1)^{-\frac{d}{2}} &  (y+1)^{-\frac{d}{2}+2}
    - (y-1)^{-\frac{d}{2}+2} + ( d-4 ) (y-1)^{-\frac{d}{2}+1} 
\\ \hline \hline
\multicolumn{3}{||c||}{\mbox{Edwards-Wilkinson {\sc ew}2}} &
    (y-1)^{-\frac{d}{2}}  & (y-1)^{-\frac{d}{2}+1+\rho} -
    (y+1)^{-\frac{d}{2}+1+\rho} \\ \hline 
\multicolumn{3}{||c||}{\mbox{Mullins-Herring {\sc mh}1}} &
    (y-1)^{-\frac{d}{4}}  & (y-1)^{-\frac{d}{4}+1} -
    (y+1)^{-\frac{d}{4}+1} \\ \hline
\multicolumn{3}{||c||}{\mbox{Mullins-Herring {\sc mh}2}} &
    (y-1)^{-\frac{d}{4}}  & (y-1)^{-\frac{d}{4}+1+\rho/2} -
    (y+1)^{-\frac{d}{4}+1+\rho/2} \\ \hline 
\multicolumn{3}{||c||}{\mbox{Mullins-Herring {\sc mh}c}} &
    (y-1)^{-{(d+2)}/{4}}  & (y-1)^{-{(d-2)}/{4}} -
    (y+1)^{-{(d-2)}/{4}} \\ \hline \hline
  \end{array}
  \]
\caption[Scaling functions]{Scaling functions (up to normalization) of the 
autoresponse and autocorrelation of the critical bosonic contact and bosonic 
pair-contact processes \cite{Baum05a}. 
The scaling functions for some simple growth models \cite{Roet06,Baum06e}, 
as defined in section~4, are also listed.}
\label{tb:results_fun}
\end{table}

\section{The bosonic processes and local scale-invariance}

We now show that the exact results for response and correlation functions
of the {\sc bcp} and the {\sc bpcp} as listed in table~\ref{tb:results_fun}
can be understood from local scale-invariance \cite{Baum05b,Baum06b}.

\subsection{Bosonic contact process}

The master equation which describes the critical bosonic contact process 
can be turned into a field-theory in a standard fashion
through an operator formalism which uses a particle annihilation operator
$a(t,\vec{r})$ and its conjugate $a^{\dag}(t,\vec{r})$ \cite{Doi76,Taeu05}. 
For calculating connected correlators, it is useful to define the
shifted fields 
\BEA
\phi(t,\vec{r}) &:=& a(t,\vec{r}) - \rho_0 \nonumber \\
\wit{\phi}(t,\vec{r}) &:=& \bar{a}(t,\vec{r}) 
= a^{\dag}(t,\vec{r}) - 1
\label{gl:phiphit}
\EEA
such that $\langle \phi(t,\vec{r}) \rangle = 0$ (our notation implies a mapping
between operators and quantum fields, using the known equivalence between 
the operator formalism and the path-integral
formulation \cite{Droz94,Howa97,Taeu05}). As we shall see, these
fields $\phi$ and $\wit{\phi}$ will become the natural quasi-primary fields
from the point of view of local scale-invariance. 
We remark that the response function is not affected by
this shift, since 
\BEQ
R(t,s;\vec{r},\vec{r}') = \left.\frac{\delta \langle a(t,\vec{r})
\rangle}{\delta h(s,\vec{r}')}\right|_{h=0} = 
\left.\frac{\delta \langle \phi(t,\vec{r})
\rangle}{\delta h(s,\vec{r}')}\right|_{h=0} 
\EEQ
As for magnets, the field-theoretical action \cite{Howa97} is again decomposed
$J[\phi,\wit{\phi}]=J_0[\phi,\wit{\phi}]+J_b[\phi,\wit{\phi}]$ into
a `deterministic' part 
\BEQ
J_0[\phi,\wit{\phi}] := \int \!\D \vec{R} \int \!\D u\:
\left[ \wit{\phi} (2\mathcal{M}\partial_u - \nabla^2)\phi \right]
\EEQ
and which is manifestly Galilei-invariant, 
whereas the `noise' is described by 
\BEQ
J_b[\phi,\wit{\phi}] := - \mu \int \!\D \vec{R} \int \!\D u\: 
\left[{\wit{\phi}}^2 (\phi+\rho_0) \right].
\EEQ
We use uncorrelated initial conditions $C(0,0;\vec{r})=0$ throughout.

In what follows, some composite fields will be needed, which we list,
together with their scaling dimensions and their masses, in 
table~\ref{tab:champs}. 
\begin{table}
\begin{center}
\begin{tabular}{|c|cr|} \hline 
field & scaling dimension & mass \\ \hline 
$\phi$       & $x$ & $\cal M$ \\
$\wit{\phi}$ & $\wit{x}$ & $-{\cal M}$ \\
${\wit{\phi}}^2$ & $\wit{x}_2$ & $-2{\cal M}$ \\
$\Upsilon := {\wit{\phi}}^2 \phi$ & $x_{\Upsilon}$ & $-{\cal M}$ \\ 
$\Sigma :=  {\wit{\phi}}^3 \phi$ & $x_{\Sigma}$ & $-2 {\cal M}$  \\ 
$\Gamma:=  {\wit{\phi}}^3 \phi^2$ & $x_{\Gamma}$ & $-{\cal M}$  \\ \hline
\end{tabular} \end{center}
\caption{Scaling dimensions and masses of some composite fields.
\label{tab:champs}}
\end{table}
We remark that for free fields one has
\BEQ
\wit{x}_2 = 2 \wit{x} \;\; , \;\;
x_{\Upsilon} = 2 \wit{x}+x\;\; , \;\;
x_{\Sigma} = 3\wit{x} + x \;\; , \;\;
x_{\Gamma} = 3\wit{x} + 2 x
\EEQ
but these relations need no longer hold for interacting fields. On the other
hand, from the Bargman superselection rules we
expect that the masses of the composite fields as given in
table~\ref{tab:champs} should remain valid for interacting fields as well.

As in section~1, we now have a similar reduction to averages of the
noiseless theory. 
First, for the computation of the response function, we add the
term $\int \!\D \vec{R} \int \D u\,  \wit{\phi}(u,\vec{R})
h(u,\vec{R})$ to the action. As usual the response function is \cite{Baum05b}
\BEA
\lefteqn{R(t,s;\vec{r},\vec{r}') = \left\langle \phi(t,\vec{r})
\wit{\phi}(s,\vec{r}') \right\rangle} \nonumber \\
&=& \left \langle \phi(t,\vec{r})
\wit{\phi}(s,\vec{r}') \exp \left( - \mu \int \!\D \vec{R}
\int \!\D u\: \wit{\phi}^2(u,\vec{R}) 
(\phi(u,\vec{R}) + \rho_0) \right) \right \rangle_0 
\nonumber \\
&=& \left\langle \phi(t,\vec{r}) \wit{\phi}(s,\vec{r}') \right\rangle_0
= R_0(t,s;\vec{r},\vec{r'})
\label{gl:Rsansbruit}
\EEA
where we expanded the exponential and applied the Bargman
superselection rule. Indeed, the two-time response is just given by the
response of the (gaussian) noise-less theory. We have therefore
reproduced the exact result of table~\ref{tb:results_fun} 
for the response function of the critical bosonic contact process.

Second, we have for the correlator
\BEA
C(t,s;\vec{r},\vec{r}') &= &\left\langle \phi(t,\vec{r})
\phi(s,\vec{r}') \exp \left( -\mu \int \!\D \vec{R} \int \!\D u\:
\wit{\phi}^2(u,\vec{R}) \phi(u,\vec{R}) \right) \right.
\nonumber \\ 
& & \left. ~~ \times \exp \left( -\mu \rho_0 \int \!\D \vec{R} \int \!\D u\:
\wit{\phi}^2(\vec{R},u) \right) \right\rangle_0
\EEA
Expanding both exponentials separately and using the Bargman superselection
rule (\ref{gl:5:Bargman}) it follows that $C=C_1+C_2$ can be written
as the sums of two terms which read 
\BEQ
\label{gl:cont1}
C_1(t,s;\vec{r},\vec{r}') = - \mu \rho_0 \int \!\D \vec{R}
\int \!\D u\: \left\langle \phi(t,\vec{r}) \phi(s,\vec{r}')
\wit{\phi}^2(u,\vec{R}) \right\rangle_0
\EEQ
and 
\BEQ
\hspace{-2.0truecm}
C_2(t,s;\vec{r},\vec{r}') = \frac{\mu^2}{2} \int \!\D
\vec{R} \D \vec{R}' \int \!\D u \D u'\: \left\langle \phi(t,\vec{r})
\phi(s,\vec{r}') \Upsilon(u,\vec{R}) \Upsilon(u',\vec{R}') \right\rangle_0
\EEQ
using the field $\Upsilon$, see table~\ref{tab:champs}. 
Hence the connected correlator is determined by three- 
and four-point functions of the noiseless theory. 

The noiseless three-point response needed for $C_1$ can be found
from its covariance under the ageing algebra \cite{Baum05b}
\BEA
\langle \phi(t,\vec{r}) \phi(s,\vec{r}') \wit{\phi}^2
(u,\vec{R}) \rangle_0 &=&
(t-s)^{x-\frac{1}{2}\tilde{x}_2} (t-u)^{-\frac{1}{2} \tilde{x}_2}
(s-u)^{-\frac{1}{2} \tilde{x}_2} \nonumber \\ & &
\hspace{-5.5truecm} \times \exp \left( -\frac{\mathcal{M}}{2}
\frac{(\vec{r}-\vec{R})^2}{t-u} - \frac{\mathcal{M}}{2}
\frac{(\vec{r}'-\vec{R})^2}{s - u} \right) \Psi_3 ( u_1,v_1)
\Theta(t-u) \Theta(s-u)
\label{gl:55}
\EEA
with
\BEA
u_1 & = & \frac{u}{t}\cdot \frac{ [(s-u)(\vec{r}-\vec{R})
- (t-u)(\vec{r}' - \vec{R})]^2 }{(t-u) (s-u)^2} \nonumber \\
v_1 & = & \frac{u}{s}\cdot \frac{ [(s-u)(\vec{r}-\vec{R})
- (t-u)(\vec{r}' - \vec{R})]^2 }{(t-u)^2 (s-u)} 
\EEA
and an undetermined scaling function $\Psi_3$. The
$\Theta$-functions express causality. Specifically, for the 
autocorrelator, i.e. $\vec{r} = \vec{r}'$ this yields, with $y=t/s$
\newpage \typeout{ *** saut de page *** }
\BEA
\lefteqn{C_1(t,s) = -\mu \rho_0\, s^{-x - \frac{1}{2}\tilde{x}_2 +
\frac{d}{2} + 1} \cdot (y-1)^{-(x - \frac{1}{2} \tilde{x}_2)}} 
\nonumber \\ 
& & \hspace{-0.5cm} \times \int_0^1 \!\D
\theta\, (y-\theta)^{-\frac{1}{2} \tilde{x}_2}
(1-\theta)^{-\frac{1}{2}
\tilde{x}_2} \int_{\mathbb{R}^d} \!\D \vec{R}\, \exp \left(
-\frac{\mathcal{M}}{2} \vec{R}^2
\frac{y+1-2\theta}{(y-\theta)(1-\theta)} \right)\nonumber
\\& & \hspace{-0.5cm} \times H \left(
\frac{\theta}{y}
\frac{\vec{R}^2 (y-1)^2}{(y-\theta)(1-\theta)^2}, \theta\,
\frac{\vec{R}^2 (y-1)^2}{(y-\theta)^2 (1-\theta)} \right)
\EEA
where $H$ is an undetermined scaling function. A similar, but
quite lengthy, expression can be 
derived for $C_2$ and depends on $\tilde{x}_2$ and $x_{\Upsilon}$ 
\cite{Baum05b}. Since the critical {\sc bcp} is described by a free
field-theory, one can expect from table~\ref{tab:expoforaequi} that 
$x=\wit{x}=d/2$ and hence for the composite fields
\BEQ
\tilde{x}_2 = d, \qquad x_{\Upsilon} = \frac{3}{2} d
\EEQ
Consequently, the autocorrelator takes the general form
\BEQ
C(t,s) = s^{1-d/2} g_1(t/s) + s^{2-d} g_2(t/s)
\EEQ
For $d$ larger than the lower critical dimension $d_*=2$, the second term
merely furnishes a finite-time correction. On the other hand, for $d<d_*=2$,
it would be the dominant one and we can only achieve agreement by discarding
the scaling function $g_2$. It remains to be seen that $g_1$ is 
compatible with the exact result given in table~\ref{tb:results_fun}. 

This can be achieved by choosing in eq.~(\ref{gl:55}) \cite{Pico04}
\BEQ \label{gl:3:Xi}
\Psi_3(u_1,v_1) = \Xi \left(\frac{1}{u_1}-\frac{1}{v_1}\right)
\EEQ
where $\Xi$ remains an arbitrary function. Then
\BEA
C_1(t,s) &=& -\mu \rho_0 s^{\frac{d}{2}+1-x-\frac{1}{2}\tilde{x}_2}
(t/s-1)^{\frac{1}{2} \tilde{x}_2 - x - \frac{d}{2}} 
\nonumber \\
& & \times \int_0^1 \!\D \theta\,
[(t/s -\theta)(1-\theta)]^{\frac{d}{2}-\frac{1}{2} \tilde{x}_2}
\phi_1 \left(\frac{t/s+1-2 \theta}{t/s-1} \right)
\EEA
where the function $\phi_1$ is defined by 
\BEQ
\phi_1(w) = \int \!\D \vec{R}\, \exp\left(
-\frac{\mathcal{M}w}{2} \vec{R}^2 \right) \Xi(\vec{R}^2)
\EEQ
Now the result for the {\sc bcp} in table~\ref{tb:results_fun} is recovered
if one chooses \cite{Pico04,Baum05b}
$\phi_1(w) = \phi_{0,c} w^{-1-a}$. 
This form for $\phi_1(w)$ guarantees
that the three-point response function 
$\langle\phi(\vec{r},t) \phi(\vec{r},s) \phi^2(\vec{r}',u) \rangle_0$ is 
nonsingular for $t = s$.\footnote{We remark that for $2<d<4$,
the same form of the autocorrelation function is also found in the critical
voter-model \cite{Dorn98}.}  

\subsection{Bosonic pair-contact process}

The construction of the action follows standard lines \cite{Howa97}. 
Making the same shift eq.~(\ref{gl:phiphit}) as before, the
action becomes $J[\phi,\wit{\phi}] =J_0[\phi,\wit{\phi}] + J_b[\phi,\wit{\phi}]$
where the `deterministic' part now reads 
\BEQ
\label{gl:sigma02}
J_0[\phi,\wit{\phi}] := \int \!\D \vec{r} \int \!\D t\:
\left[ \wit{\phi} (2\mathcal{M}\partial_t - \nabla^2)\phi - \alpha\wit{\phi}^2
\phi^2\right].
\EEQ
The remaining part is the noise-term which reads
\BEA
\label{gl:kappa2}
\hspace{-1.8cm}
J_b[\phi,\wit{\phi}] & = & \int \!\D \vec{R} \int \!\D u\:
\left[- \alpha \rho_0^2 \wit{\phi}^2 - 2 \alpha \rho_0\wit{\phi}^2  \phi
-\mu \wit{\phi}^3 \phi^2- 2 \mu \rho_0
\wit{\phi}^3 \phi - \rho_0^2 \wit{\phi}^3 \right]
\EEA
The discussion of the Schr\"odinger- or, more precisely, the
ageing-invariance of $J_0$ can no longer use
the representations we considered so far, since the equation of motion
associated to $J_0$ is non-linear, viz. 
\BEQ \label{gl:NLS}
2\mathcal{M}\partial_t \phi(\vec{x},t) = \nabla^2 \phi(\vec{x},t) 
-g \phi^2(\vec{x},t) \wit{\phi}(\vec{x},t)
\EEQ
While for a constant $g$ the well-known symmetries of this equation are those
encountered before, 
it was pointed out recently that $g$ rather should be considered as a 
dimensionful quantity and hence should transform under 
local scale-transformations as well \cite{Stoi05}. The requires an extension
of the generators of $\mathfrak{age}_d$ which do contain a dimensionful 
coupling $g$ \cite{Baum05b}. Then it can be shown that {\em the
Bargman superselection rules (\ref{gl:5:Bargman}) still apply} and
the response function of the noiseless theory now reads \cite{Baum05b} 
\BEA
R_0(t,s;\vec{r},\vec{r}') &=& (t-s)^{-\frac{1}{2} (x_1 + x_2)} \left(
\frac{t}{s} \right)^{-\frac{1}{2} (x_1 - x_2)} 
\nonumber \\
& &\times \exp\left( -\frac{\mathcal{M}}{2} \frac{(\vec{r}-\vec{r}')^2}{t-s}
\right) \tilde{\Psi}_2 \left(\frac{t}{s} \cdot \frac{t-s}{g^{1/y}},
\frac{g}{(t-s)^y} \right)
\EEA
with an undetermined scaling function $\tilde{\Psi}_2$. In these calculation, 
we have assumed for technical simplicity that each field $\vph_i$ has a 
coupling $g_i$ and only at the end, we let $g_1 = \ldots = g_n = g$. This
form is clearly consistent with the results for the {\sc bpcp} listed 
in table~\ref{tb:results_fun}, 
for both $\alpha<\alpha_c$ and $\alpha=\alpha_c$, if we identify
\BEQ
x := x_1 = x_2 = a+1 = \frac{d}{2} \;\; , \;\;
\tilde{\Psi}_2 = \mbox{\rm const.}
\EEQ
In distinction with the bosonic contact process, the symmetries of the
noiseless part $S_0$ do not fix the response function completely but leave a 
certain degree of flexibility in form of the scaling function
$\tilde{\Psi}_2$.  

As before, averages can be reduced to averages within the
`deterministic' theory only. Technically, calculations become a little
longer for the {\sc bpcp}, since because of the structure of $J_b$ several
composite field must be defined. We refer to \cite{Baum05b} for the details
and merely quote here the results. First, the response function
does not depend explicitly on the noise, viz. 
\BEQ \label{gl:4:R}
R(t,s;\vec{r},\vec{r}') = R_0(t,s;\vec{r},\vec{r}')
\EEQ
Second, the results of table~\ref{tb:results_fun} for $f_C(y)$ 
can be reproduced from the single term 
\BEQ \label{gl:4:G1}
C_1(t,s) = \alpha \rho_0^2 \int \!\D \vec{R} \int \!\D u\: 
\left\langle
\phi(t,\vec{r}) \phi(s,\vec{r}) \wit{\phi}^2(u,\vec{R})
\right\rangle_0
\EEQ
The required three-point function now reads 
\BEA
\left\langle \phi(t,\vec{r}) \phi(s,\vec{r}') \tilde{\phi}^2(u,\vec{R})
\right\rangle_0 &=&
(t-s)^{x-\frac{1}{2} \tilde{x}_2} (t-u)^{-\frac{1}{2} \tilde{x}_2}
(s-u)^{-\frac{1}{2} \tilde{x}_2} \nonumber \\ & &
\hspace{-4.0truecm} \times \exp \left( -\frac{\mathcal{M}}{2}
\frac{(\vec{r}-\vec{R})^2}{t-u} - \frac{\mathcal{M}}{2}
\frac{(\vec{r}'-\vec{R})^2}{s - u} \right) \tilde{\Psi}_3 (
u_1,v_1,\beta_1,\beta_2,\beta_3)
\EEA
with
\BEA
u_1 & = & \frac{u}{t}\cdot \frac{ [(s-u)(\vec{r}-\vec{R})
- (t-u)(\vec{r}' - \vec{R})]^2 }{(t-u) (s-u)^2} \\
v_1 & = & \frac{u}{s}\cdot \frac{ [(s-u)(\vec{r}-\vec{R})
- (t-u)(\vec{r}' - \vec{R})]^2 }{(t-u)^2 (s-u)}  \\
\beta_1 & = & \frac{1}{s_2} \cdot \frac{\alpha^{1/y}}{(t-u)^2},
\; \; \beta_2  =  \frac{1}{s_2} \cdot \frac{\alpha^{1/y}}{(s-u)^2},
\;\; \beta_3  =  \alpha^{1/y} s_2 \\
s_2 & = & \frac{1}{t-u} + \frac{1}{u}
\EEA
Next, we choose the following realization for $\tilde{\Psi}_3$
\BEQ
\label{gl:some_equation}
\tilde{\Psi}_3 (u_1,v_1,\beta_1,\beta_2,\beta_3) = \Xi \left(
\frac{1}{u_1} - \frac{1}{v_1} \right) \left[ -
\frac{(\sqrt{\beta_1}-\sqrt{\beta_2}) \sqrt{\beta_3}
}{\beta_3 - \sqrt{\beta_2 \beta_3} } \right]^{(a-b)}
\EEQ
where the scaling function $\Xi$ was already encountered in 
eq.~(\ref{gl:3:Xi}) for the bosonic contact process. We now have to distinguish
the two different cases $\alpha < \alpha_C$ and $\alpha = \alpha_C$. 
For the first case $\alpha<\alpha_C$, we have
$a-b = 0$ so that the last factor in
(\ref{gl:some_equation}) disappears and we simply return to
the expressions already found for the bosonic contact process, in agreement
with the known exact results. However, at the multicritical point
$\alpha=\alpha_C$ we have $a-b \neq 0$ and the last factor becomes
important. We point out that only the  presence or absence of this factor 
distinguishes the cases $\alpha < \alpha_C$ and $\alpha =
\alpha_C$. 

Inserting the values of the $\beta_{1,2,3}$ we finally obtain for the
autocorrelation function
\BEA
C_1(t,s) &=& s^{-b} (y-1)^{(b-a)-a-1} \int_0^1 \!\D \theta\,
[(t/s-\theta)(1-\theta)]^{a-b}\nonumber \\ 
& & \times \phi_1\left( \frac{t/s+1-2 \theta}{t/s-1}
\right) \left[ \frac{\theta (t/s-1)}{(t/s-\theta) (1-\theta)}
\right]^{a-b} 
\EEA
where we have identified
\BEQ
\tilde{x}_2 = 2(b-a) + d
\EEQ
$C_1(t,s)$ reduces to the expression (\ref{gl:4:bcp:formaintegralC}) if we
choose the same  $\phi_1(w)=\phi_{0,c} w^{-1-a}$ as before. 
Hence all scaling functions for the {\sc bpcp} are reproduced correctly.

\section{Growth models}

As a further illustration on a different class of system we now describe
recent work by R\"othlein, Baumann and Pleimling \cite{Roet06} on kinetic growth
models. As a simple model for ballistic deposition consider the Family
model \cite{Fami86}. A particle is randomly dropped onto the sites of a lattice. 
However, before it is fixed, the particle explores the sites around the one it
arrived at (typically the nearest neighbours) and fixes itself at the lattice
site with the lowest height. One obtains in this way a growing surface which
may be described in terms of a height variable $h(t,\vec{r})$. Clearly, 
since there
is only irreversible deposition, the system will never arrive at an 
equilibrium state. 

Some of the simplest continuum descriptions of these phenomena can be cast
into stochastic linear equations for the height variable. For the sake
of simplicity, we shall always work in the frame co-moving with the
mean surface height. For example, if the
deposition is purely diffusive and without mass conservation, the simplest model
is the well-known Edwards-Wilkinson ({\sc ew}) model \cite{Edwa82}
\BEQ \label{gl:EW}
\partial_t h(t,\vec{r}) = D \nabla^2 h(t,\vec{r}) +\eta(t,\vec{r})
\EEQ
but on the other side, if mass conservation must be taken into account, one 
might rather consider the Mullins-Herring ({\sc mh}) model, see \cite{Wolf90}
\BEQ \label{gl:MH}
\partial_t h(t,\vec{r}) = -D (\nabla^2)^2 h(t,\vec{r}) + \eta(t,\vec{r})
\EEQ
Following \cite{Roet06}, the following types of gaussian noise with
vanishing first moment $\langle \eta(t,\vec{r})\rangle=0$ will be
considered:
\begin{enumerate}
\item[(a)] non-conserved, short-ranged
$\langle \eta(t,\vec{r})\eta(s,\vec{r}')\rangle = 2D
\delta(\vec{r}-\vec{r}')\delta(t-s)$.
\item[(b)] non-conserved, long-range
$\langle \eta(t,\vec{r})\eta(s,\vec{r}')\rangle = 2D
|\vec{r}-\vec{r}'|^{2\rho-d}\delta(t-s)$ and $0<\rho<d/2$.
\item[(c)] conserved, short-ranged
$\langle \eta(t,\vec{r})\eta(s,\vec{r}')\rangle = -2D
\nabla_{\vec{r}}^2 \delta(\vec{r}-\vec{r}')\delta(t-s)$.
\end{enumerate}
Then the following models were studied in \cite{Roet06}:
\begin{enumerate}
\item {\sc ew}1: eq.~(\ref{gl:EW}) with the non-conserved noise (a). 
\item {\sc ew}2: eq.~(\ref{gl:EW}) with the non-conserved, long-ranged noise (b).
\item {\sc mh}1: eq.~(\ref{gl:MH}) with the non-conserved noise (a).
\item {\sc mh}2: eq.~(\ref{gl:MH}) with the non-conserved, long-ranged noise (b).
\item {\sc mh}c: eq.~(\ref{gl:MH}) with the conserved noise (c) \cite{Baum06e}.
\end{enumerate}
In the models {\sc ew}1 and {\sc mh}c the noise is in agreement with detailed
balance while for the other models it is not. Solving the linear equations
(\ref{gl:EW}) and (\ref{gl:MH}) is straightforward. The two-time correlation and
response functions are seen to obey the same kind of scaling behaviour as for
the non-equilibrium models considered before. In table~\ref{tab:expoforaequi}
the values of the exponents are listed and the scaling functions for the
autoresponse and the autocorrelation are included in table~\ref{tb:results_fun}
(the models {\sc bcp} and {\sc ew}1 lead to identical results and are not 
listed separately). The result quoted for the {\sc mh}c model is only valid 
for $d>2$ as stated; for $d=2$ the scaling function becomes
$f_C(y)=2D \ln[(y-1)/(y+1)]$ \cite{Baum06e}. Detailed simulations show that 
the correlation and response functions of the Family model \cite{Fami86} and 
of a variant of it are perfectly
described by the {\sc ew}1 model and hence should be in the same universality
class \cite{Roet06}. 

We shall see shortly that all results about the scaling functions as listed in 
table~\ref{tb:results_fun} can be understood from an extension of dynamical scaling 
to local scale-invariance, for both the {\sc ew} models, where the dynamical
exponent $z=2$, as well as for the {\sc mh} models, where $z=4$. In this
context, it is instructive to consider the space-time responses 
$R(t,s;\vec{r})$ as well. In general, one finds the structure
\BEQ \label{Resptem}
R(t,s;\vec{r}) = R(t,s) \Phi(|\vec{r}| (t-s)^{-1/z})
\EEQ
where $R(t,s)=R(t,s;\vec{0})$ is the autoresponse function. If $z=2$, one has a simple exponential form $\Phi(u)=\exp(-{\cal M}u)$ \cite{Henk94}\footnote{For
detailed quantitative tests in Ising and Potts models, 
see \cite{Henk03b,Lore06}.} while for $z=4$, the form
of $\Phi(u)$ is more complicated. For example, in the {\sc mh}c model with conserved noise \cite{Baum06e}  
\BEQ
\label{result_response}\label{gl:Rmhc}
\hspace{-1truecm} \Phi(u) = 
\Phi_0 \left[{_0F_2}\left(
\frac{1}{2},\frac{d}{4};\frac{u^4}{256}\right) - 
\frac{8}{d}\frac{\Gamma\left(\frac{d}{4}+1\right)}{\Gamma
\left(\frac{d}{4}+\frac{1}{2}\right)}
\sqrt{\frac{u^4}{256}\,}\;
{_0F_2}\left(\frac{3}{2},\frac{d}{4}+\frac{1}{2};
\frac{u^4}{256}\right)\right]
\EEQ
where $\Phi_0$ is a known constant and ${}_0F_2$ is a hypergeometric function. 
Very similar expressions have been derived in the other {\sc mh} models, see
\cite{Roet06} for details. 

Understanding the form of the spatio-temporal response allows for an explicit test of one of the important ingredients of the theory of local scale-invariance, namely Galilei-invariance for $z=2$ and its generalization if
$z\ne 2$. Since in the {\sc ew} models one has $z=2$, the calculation of
$R(t,s;\vec{r})$ is a direct extension of the discussion presented in section~1.
Since this has been discussed in detail in the litterature \cite{Henk94,Henk02}
we shall not repeat it here and rather concentrate on the case $z=4$, following
\cite{Henk02,Roet06,Baum06e}. 

Consider the dynamical symmetries of the `Schr\"odinger operator'
\BEQ
\label{diff_operator}
{\cal S}_4 := -\lambda \partial_t + \frac{1}{16}
\left(\nabla_{\vec{r}}^2\right)^2.
\EEQ
which will become related to the deterministic part of the Mullins-Herring
equation. As before, we ask if dynamical scaling can be extended to a larger set of local scale-transformations, given standard dynamical scaling and spatial translation-invariance. Specializing the construction of \cite{Henk02} 
to $z=4$, these generators read as follows, with 
the shorthands $\vec{r} \cdot \partial_{\vec{r}} := \sum_{k=1}^d r_k\,
\partial_{r_k}$,$\nabla_{\vec{r}}^2 := \sum_{k=1}^d\partial_{r_k}^2 $ and 
$\vec{r}^2 := \sum_{k=1}^d r_k^2$, 
\newpage \typeout{ *** saut de page ***}
\BEA
X_{-1} & := & -\partial_t  \nonumber \\
X_0 & := & -t \partial_t - \frac{1}{4}\, \vec{r} \cdot
\partial_{\vec{r}} - \frac{x}{4} \nonumber \\
X_1 & := & - t^2 \partial_t - \frac{
x}{2} t - \lambda \vec{r}^2 \,
(\nabla_{\vec{r}}^2)^{-1}
- \frac{1}{2} \, t \vec{r} \cdot \partial_{\vec{r}}
\nonumber \\ 
& & +4 \gamma \, \left( \vec{r}\cdot \partial_{\vec{r}}
\right) \, \left( \nabla_{\vec{r}}^2 \right)^{-2}  +
2\gamma (d-4) \left( \nabla_{\vec{r}}^2 \right)^{-2}
\label{gl:alg4} \\
R^{(i,j)} &:=& r_i \partial_{r_j} - r_j \partial_{r_i}
\;\; ; \;\; \hspace{4truecm} \mbox{\rm where $1\leq i < j\leq d$} 
\nonumber \\
Y^{(i)}_{-1/4} & = & - \partial_{r_i} \nonumber \\
Y^{(i)}_{3/4} & = & - t \partial_{r_i} - 4 \lambda
r_i \left( \nabla_{\vec{r}}^2
\right)^{-1} + 8 \gamma \, 
\partial_{r_i} \left( \nabla_{\vec{r}}^2 \right)^{-2} \nonumber
\EEA
where $x$ is the scaling dimension of the fields on which these
generators act and $\gamma,\lambda$ are further field-dependent parameters.
Here, the generators $X_{\pm 1,0}$ correspond to projective changes in 
the time $t\mapsto (\alpha t+\beta)/(\gamma t+\delta)$ with $\alpha\delta-\beta\gamma=1$, the generators $Y^{(i)}_{n-1/4}$ are 
space-translations, generalized Galilei-transformations and so on 
and $R^{(i,j)}$ are spatial rotations. Here we use the following properties
$\partial_r^{\alpha} \partial_r^{\beta} = \partial_r^{\alpha+\beta}$ and
$\left[ \partial_r^{\alpha}, r \right] = \alpha \partial_r^{\alpha-1}$, which can be proven for fractional derivatives with extra distributional terms \cite[Appendix A]{Henk02} which in turn are closely
related to fractional derivatives as defined in \cite{Gelf64}. 
Furthermore, $(\nabla_{\vec{r}}^2)^{-2} =
(\nabla_{\vec{r}}^2)^{-1} \cdot (\nabla_{\vec{r}}^2)^{-1}$ and the operator 
$(\nabla_{\vec{r}}^2)^{-1}$ is defined, e.g., for $d=2$, by formal expansion 
\cite{Roet06,Baum06e} 
\BEQ
(\nabla_{\vec{r}}^2)^{-1} := (\partial_{r_1}^2 +
\partial_{r_2}^2)^{-1} := \sum_{n=0}^\infty (-1)^n
\partial_{r_1}^{-2-2n} \partial_{r_2}^{2 n}
\EEQ
This implies the commutator 
${[}(\nabla_{\vec{r}}^2)^n,r_i] = n\, \partial_{r_i}
(\nabla_{\vec{r}}^2)^{n-1}$ for all $n\in\mathbb{Z}.$  

That the generators eq.~(\ref{gl:alg4}) indeed describe dynamical symmetries 
of the `Schr\"odinger operator' (\ref{diff_operator}) now follows from the
commutators \cite{Henk02,Baum06e}
\BEQ
{} [{\cal S}_4,Y^{(i)}_{-1/4}] \:=\: [{\cal S}_4,Y^{(i)}_{3/4}] 
\:=\: [{\cal S}_4,R^{(i,j)}] \:=\: 0 \;\; , \;\;
{} [{\cal S}_4,X_0] = - {\cal S}_4 
\EEQ
This means that for a solution of the `Schr\"odinger equation' 
${\cal S}_4\phi=0$ the transformed function ${\cal X}\phi$ is again solution of
the `Schr\"odinger equation'. Finally 
\BEQ
{}   [{\cal S}_4,X_1] = -2 t {\cal S}_4 + \frac{\lambda}{2} 
   \left( x - \frac{d}{2} -1 +\frac{2\gamma}{\lambda} \right) 
\EEQ
hence a dynamical symmetry is obtained if the field $\phi$ has the scaling
dimension
\BEQ
\label{rel_scaling_dim}
x = \frac{d}{2} + 1-\frac{2 \gamma}{\lambda}
\EEQ
Generalizing from conformal or Schr\"odinger-invariance, quasiprimary
fields transform covariantly under the generators (\ref{gl:alg4}) and 
in particular the response function will satisfy the conditions
$X_1 R = X_0 R = Y_{-1/4}^{(i)}R = R^{(i,j)}R =0$ (the other conditions
then follow from the Jacobi identities) and
is now characterized by its scaling dimension $x_i$ and the further
parameters $\gamma_i,\lambda_i$. 
In calculating the response function $R=\langle \phi \wit{\phi}\rangle$ 
this leads to the conditions $\lambda=-\wit{\lambda}$ and $\gamma=-\wit{\gamma}$
whereas the scaling function $\Phi(u)$ from eq.~(\ref{Resptem}) 
can be found by solving the differential equation 
\BEA
\label{scaling_function}
\hspace{-1.5cm}
\left(\partial_u \left( \frac{1}{u^{d-1}} \partial_u \left(
u^{d-1} \partial_u \right) \right)^2 + 4\lambda u \left(
\frac{1}{u^{d-1}} \partial_u \left( u^{d-1} \partial_u
\right) \right) - 16 \gamma \partial_u\right) \overline{\Phi}(u) & = &
0
\EEA
where
\BEQ
\label{sc1:sc2}
{\Phi}(u) = \left( \frac{1}{u^{d-1}} \partial_u
\left(u^{d-1} \partial_u \right) \right)^{2} \overline{\Phi}(u)
\EEQ
Solving this via series expansion techniques \cite{Henk02,Roet06,Baum06e} and checking carefully that
all independent solutions are taken into account, one can indeed recover the explicit result (\ref{gl:Rmhc}) for the {\sc mh}c model 
as a special case \cite{Baum06e} and, similarly, also for the {\sc mh}1 and {\sc mh}2 models \cite{Roet06}. 

Lastly, since the {\sc mh} equation is linear, there is a natural Wick theorem
which allows to go over from the stochastic Langevin equation to the
deterministic equation in quite an analogous way as previously for $z=2$
\cite{Roet06,Baum06e}. The extension of the technique to non-linear cases and/or to $z\ne 2,4$ is work in progress and will be reported elsewhere \cite{Baum06f}. In a similar
way one may also check that the correlation functions agree with LSI. 

The {\sc mh} models considered in this section are, together with the
critical spherical model with a conserved order-parameter \cite{Baum06e}, 
the first analytically 
solved examples with $z\ne 2$ where local scale-invariance could be fully 
confirmed. These examples make it in particular clear that the height of the
surface in growth processes is a natural candidate for being described by a 
quasiprimary scaling operator of local scale-invariance.  

\section{Conclusions}

In this survey, we have reviewed to what extent one may expect that a 
phenomenological description, which has been successfully applied to describe 
the ageing of
magnetic systems relaxing towards equilibrium steady-states, may be extended
to more general models where the stationary states are no longer part of an
equilibrium statistical ensemble. This situation frequently arises in chemical kinetics, see \cite{Alca94,Argy01,Copp04,Voit05a,Voit05b}, and the 
consideration of such systems is of interest in studies of ageing in 
chemical/biological systems where 
already the intrinsic microscopic dynamics and/or constraints does not 
admit relaxation 
towards thermal equilibrium. It is clear that the study of ageing phenomena
without detailed balance still stands at its very beginning and many open 
questions remain. In particular, the few models reviewed here certainly do not exhaust all possibilities for ageing behaviour without detailed balance but should be rather seen as case studies whose results might suggest further research problems. This review has already served a useful purpose if it
encourages people to explore more systematically the properties of two-time
observables of non-equilibrium systems, e.g. reaction-diffusion problems
or biologically motivated models. 

Specifically, the following points should be noted:
\begin{enumerate}
\item the generic scaling form (\ref{gl:cpc:ablCR}) was seen to be satisfied 
in all models considered. However, the exponent relation $a=b$, known to 
hold for critical systems with detailed balance, is no longer valid in general, 
see table~\ref{tab:expoforaequi}. For different universality classes, the
relation $a=b$ is either maintained or else broken in different ways. 
This means that there is no obvious and general 
analogue of an universal limit fluctuation-dissipation ratio $X_{\infty}$
(for magnets one sometimes tries to relate this to a non-equilibrium 
temperature, see e.g. \cite{Cugl02}) even if such an analogy may be defined for certain subclasses. 
\item for the uncorrelated initial states which have been considered so far,
one observes that the autocorrelation and autoresponse exponents agree
$\lambda_C=\lambda_R$. It would be interesting to see if spatial or 
temporal disorder in the rates may change that conclusion, as it apparently
happens in diluted magnets \cite{Sche03,Henk06b}. 
\item one of our main question with respect to ageing systems has been if
dynamical scaling permits an extension to a larger group of local 
scale-transformation \cite{Henk02}. 
It has turned out that from the point of view of 
LSI the responses are the most easy quantities to study. The long list
of examples, see tables~\ref{Tabelle1} and \ref{Tabelle2}, where the two-time 
autoresponse function $R(t,s)$ was concluded to be in agreement with LSI is
clear evidence that LSI is indeed a successful phenomenological scheme, and
this for values of the dynamical exponent $z$ which are often far from 
$z=2$ characterizing simple diffusive motion. But this ansatz remains to be
proven, especially for $z\ne 2$, e.g. from some underlying stochastic
Langevin equation. The exactly solvable examples we have treated suggest 
that the idea of splitting the Langevin equation in a `deterministic' part with 
possible non-trivial dynamical symmetries and a `noise' part which breaks 
those can be 
taken over from magnets to more general reaction-diffusion type system,
although the noise terms can be considerably more general. 

The examples studied here also suggest that the basic physical variables of these models, such as the particle-density or the height of the surface, should be directly relatable to the quasiprimary scaling operators of LSI. 
\end{enumerate} 
It appears to us that it should be promising to investigate more systematically
the foundations and consequences of a hitherto unsuspected non-trivial
dynamical symmetry in scale-invariant non-equilibrium dynamics. 

\newpage
\noindent 
{\bf Acknowledgements:} 

It is a pleasure to thank F. Baumann, T. Enss, A. Picone, M. Pleimling, 
J.J. Ramasco, J. Richert, U. Schollw\"ock, M.A. Santos, C.A. da Silva Santos 
and S. Stoimenov for the fruitful collaborations which lead to the results 
reviewed here. I thank M. Pleimling for providing figures 1 and 2 and G. \'Odor 
for providing figure 7. 
I also thank the INFN Firenze for warm hospitality, where this work was started.




\begin{thebibliography}{999}
\bibitem{Abri04a} S. Abriet and D. Karevski, Eur. Phys. J. {\bf B37}, 47 (2004).
\bibitem{Abri04b} S. Abriet and D. Karevski, Eur. Phys. J. {\bf B41}, 79 (2004).
\bibitem{Alba01a} E.V. Albano and M.A. Mu\~noz, Phys. Rev. {\bf E63}, 031104
(2001).
\bibitem{Alca94} F.C. Alcaraz, M. Droz, M. Henkel and V. Rittenberg, Ann.
of Phys. {\bf 230}, 250 (1994). 
\bibitem{Anni06} A. Annibale and P. Sollich, J. Phys. {\bf A39}, 2853 (2006).
\bibitem{Argy01} P. Argyrakis, S.F. Burlatsky, E. Cl\'ement and G. Oshanin, Phys. Rev. {\bf E63}, 021110 (2001). 
\bibitem{Barg54} V. Bargman, Ann. of Math. {\bf 56}, 1 (1954).
\bibitem{Baum05a} F. Baumann, M. Henkel, M. Pleimling and J. Richert, 
J. Phys. A: Math. Gen. {\bf 38} 6623, (2005).     
\bibitem{Baum05b} F. Baumann, S. Stoimenov and M. Henkel, J. Phys. {\bf A39}, 
4095 (2006). 
\bibitem{Baum06a} F. Baumann and M. Pleimling, J. Phys. {\bf A39}, 1981 (2006).
\bibitem{Baum06b} F. Baumann, J. Phys. Conf. Series {\bf 40}, 86 (2006). 
\bibitem{Baum06c} F. Baumann, {\tt private communication (2006)}. 
\bibitem{Baum06d} F. Baumann and A. Gambassi, {\tt cond-mat/0610260}. 
\bibitem{Baum06e} F. Baumann and M. Henkel, {\tt cond-mat/0611652}. 
\bibitem{Baum06f} F. Baumann and M. Henkel, {\tt in preparation}. 
\bibitem{Bela84} A.A. Belavin, A.M. Polyakov and A.B. Zamolodchikov, Nucl.
Phys. {\bf B241}, 333 (1984). 
\bibitem{Bert99} L. Berthier, J.L. Barrat and J. Kurchan, Eur. Phys. J.
{\bf B11}, 635 (1999).
\bibitem{Bert01} L. Berthier, P.C.W. Holdsworth and M. Sellitto, J. Phys.
{\bf A34}, 1805 (2001). 
\bibitem{Bray94} A.J. Bray, Adv. Phys. {\bf 43}, 357 (1994). 
\bibitem{Bray94b} A.J. Bray and A.D. Rutenberg, Phys. Rev. {\bf E49}, 
R27 (1994); {\bf E51}, 5499 (1995). 
\bibitem{Cala01} P. Calabrese and A. Gambassi, Phys. Rev. {\bf E65}, 066120 (2002).
\bibitem{Cala02} P. Calabrese and A. Gambassi, Phys. Rev. {\bf E66}, 066101 (2002).
\bibitem{Cala02a} P. Calabrese and A. Gambassi, Phys. Rev. {\bf B66}, 212407 (2002).
\bibitem{Cala05} P. Calabrese and A. Gambassi, J. Phys. {\bf A38}, R181 (2005). \bibitem{Cala06} P. Calabrese, A. Gambassi and F. Krzakala, J. Stat Mech. Theor. Exp. P06016 (2006). 
\bibitem{Cane05} L. Canet, H. Chat\'e, B. Delamotte, I. Dornic and 
M. A. Mu\~noz, Phys. Rev. Lett. {\bf 95}, 1006001 (2005).
\bibitem{Cann01} S.A. Cannas, D.A. Stariolo and F.A. Tamarit, Physica 
{\bf A294}, 362 (2001). 
\bibitem{Card90} J.L. Cardy in E. Br\'ezin and J. Zinn-Justin (eds), 
{\it Fields, strings and critical phenomena}, Les Houches XLIX, North
Holland  (Amsterdam 1990).
\bibitem{Card96} J. L. Cardy and U. C. T\"auber,
Phys. Rev. Lett. {\bf 77}, 4780 (1996).
\bibitem{Cate00} M.E. Cates and M.R. Evans (eds)
{\it Soft and fragile matter}, IOP Press (Bristol 2000).
\bibitem{Copp04} M. Coppey, O. B\'enichou, J. Klafter, M. Moreau and G. Oshanin,
Phys. Rev. {\bf E69}, 036115 (2004). 
\bibitem{Cris03} A. Crisanti and F. Ritort, J. Phys. {\bf A36}, R181 (2003)
\bibitem{Cugl02} L.F. Cugliandolo, in {\it Slow Relaxation and
non equilibrium dynamics in condensed matter}, Les Houches Session 77 July 2002,
J-L Barrat, J Dalibard, J Kurchan, M V Feigel'man eds (Springer, 2003); 
{\tt cond-mat/0210312}.
\bibitem{Cugl94b} L.F. Cugliandolo, J. Kurchan, and G. Parisi, J. Physique
{\bf I4}, 1641 (1994).
\bibitem{deDo78} C. de Dominicis and L. Peliti, Phys. Rev. {\bf B18}, 
353 (1978).
\bibitem{Doi76} M. Doi, J. Phys. A: Math. Gen. {\bf 9}, 1465 and 1479 (1976).
\bibitem{Dorn98} I. Dornic, th\`ese de doctorat, Nice et Saclay 1998. 
\bibitem{Dorn01a} I. Dornic, H. Chat\'e, J. Chave and H. Hinrichsen, Phys. Rev. Lett. {\bf 87}, 5701 (2001). 
\bibitem{Droz94} M. Droz and A. McKane, J. Phys. A: Math. Gen. {\bf 27}, L467 (1994). 
\bibitem{Edwa82} S.F. Edwards and D.R. Wilkinson, Proc. Roy. Soc. London Ser. 
{\bf A381}, 17 (1982). 
\bibitem{Enss04} T. Enss, M. Henkel, A. Picone and U. Schollw\"ock,
J. Phys. {\bf A37}, 10479 (2004).
\bibitem{Enss06} T. Enss and M. Henkel, {\tt in preparation}. 
\bibitem{Evan05} M.R. Evans and T. Hanney, J. Phys. {\bf A38}, R195 (2005).
\bibitem{Fami86} F. Family, J. Phys. {\bf A19}, L441 (1986).  
\bibitem{Fedo06} A.A. Fedorenko and S. Trimper, Europhys. Lett. {\bf 74},
89 (2006). 
\bibitem{Fish88} D.S. Fisher and D.A. Huse, Phys. Rev. {\bf B38}, 373 (1988).
\bibitem{Gamb06} A. Gambassi, J. Phys. Conf. Series {\bf 40}, 13 (2006). 
\bibitem{Gelf64} I.M. Gelfand and G.E. Shilov, {\it Generalized functions}, vol. I, Academic Press (London 1964). 
\bibitem{Glau63} R.J. Glauber, J. Math. Phys. {\bf 4}, 294 (1963). 
\bibitem{Godr00a} C. Godr\`eche and J.M. Luck, J. Phys. {\bf A33}, 1151 (2000).
\bibitem{Godr00b} C. Godr\`eche and J.-M. Luck, J. Phys. {\bf A33},
9141 (2000).
\bibitem{Godr02} C. Godr\`eche and J.M. Luck, J. Phys. Cond. Matt. {\bf 14},
1589 (2002).
\bibitem{Godr06} C. Godr\`eche, in \cite{Henk06} ({\tt cond-mat/0604276}). 
\bibitem{Gras84} P. Grassberger, F. Krause and T. von der
Twer, J. Phys. {\bf A17}, L105 (1984).
\bibitem{Hage72} C.R. Hagen, Phys. Rev. {\bf D5}, 377 (1972). 
\bibitem{Henk94} M. Henkel, J. Stat. Phys. {\bf 75}, 1023 (1994). 
\bibitem{Henk01} M. Henkel, M. Pleimling, C. Godr\`eche and J.-M. Luck,
Phys. Rev. Lett. {\bf 87}, 265701 (2001).
\bibitem{Henk02} M. Henkel, Nucl. Phys. {\bf B641}, 405 (2002).
\bibitem{Henk02a} M. Henkel, M. Paessens and M. Pleimling, 
Europhys. Lett. {\bf 62}, 644 (2003)
\bibitem{Henk03} M. Henkel and J. Unterberger, Nucl. Phys. {\bf B660}, 
407 (2003). 
\bibitem{Henk03b} M. Henkel and M. Pleimling, Phys. Rev. {\bf E68}, 
065101(R) (2003). 
\bibitem{Henk03d} M. Henkel and G.M. Sch\"utz, J. Phys. {\bf A37}, 591 (2004).
\bibitem{Henk03e} M. Henkel, M. Paessens and M. Pleimling,
Phys. Rev. {\bf E69}, 056109 (2004).
\bibitem{Henk03f} M. Henkel, in A. Kundu (ed) {\it Classical and quantum nonlinear integrable systems}, ch. 10, IOP Press (Bristol 2003). 
\bibitem{Henk04} M. Henkel, Adv. Solid State Phys. {\bf 44}, 389 (2004).
\bibitem{Henk04a} M. Henkel and M. Pleimling, Europhys. Lett. {\bf 69}, 
524 (2005). 
\bibitem{Henk04b} M. Henkel, A. Picone and M. Pleimling, Europhys. Lett. 
{\bf 68}, 191 (2004).   
\bibitem{Henk05} M. Henkel, in P.L. Garrido, J. Marro and M.A. Mu\~noz (eds)
{\it Modelling cooperative behaviour in the social sciences}, AIP Conf. 
Proc. {\bf 779}, New York (2005), p. 171 (also available at {\tt cond-mat/0503739}). 
\bibitem{Henk05a} M. Henkel and M. Pleimling, J. Phys. Cond. Matt. {\bf  17},
S1899 (2005). 
\bibitem{Henk06} M. Henkel, M.  Pleimling and R. Sanctuary (eds), {\it
Ageing and the glass transition}, Springer Lecture Notes in Physics, 
Springer (Heidelberg  2007).  
\bibitem{Henk06a} M. Henkel, T. Enss and M. Pleimling, J. Phys. {\bf A39}, L589
(2006). 
\bibitem{Henk06b} M. Henkel and M. Pleimling, Europhys. Lett. {\bf 76}, 561
(2006).  
\bibitem{Hinr98} H. Hinrichsen and H.M. Koduvely, Eur. Phys. J. {\bf B5}, 257
(1998).
\bibitem{Hinr00} H. Hinrichsen, Adv. Phys. {\bf 49}, 815 (2000). 
\bibitem{Hinr06} H. Hinrichsen, J. Stat. Mech. L06001 (2006). 
\bibitem{Hohe77} P. Hohenberg and B.I. Halperin, Rev. Mod. Phys. {\bf 49}, 
435 (1977). 
\bibitem{Houc02} B. Houchmandzadeh, Phys. Rev. {\bf E66}, 052902 (2002). 
\bibitem{Howa97} M. Howard and U.C. T\"auber, J. Phys. {\bf A30}, 7721 (1997).
\bibitem{Huse89} D.A. Huse, Phys. Rev. {\bf B40}, 304 (1989). 
\bibitem{Jank06} W. Janke, in \cite{Henk06}. 
\bibitem{Jans89} H.K. Janssen, B. Schaub and B. Schmittmann, Z. Phys. {\bf B73}, 539 (1989). 
\bibitem{Jans92} H.K. Janssen, in G. Gy\"orgyi et {\em al.} (eds) {\it
{}From Phase transitions to Chaos}, World Scientific (Singapour 1992), p. 68
\bibitem{Kast68} H.A. Kastrup, Nucl. Phys. {\bf B7}, 545 (1968).
\bibitem{Kawa04} N. Kawashima and H. Rieger, in 
{\it  Frustrated magnetic systems}, H. Diep (ed.) (World Scientific, 2004).
\bibitem{Lipp00} E. Lippiello and M. Zanetti, Phys. Rev. {\bf E61}, 
3369 (2000).
\bibitem{Lore06} E. Lorenz and W. Janke, Europhys. Lett. at the press, see
{\tt http://www.physik.uni-leipzig.de/}\~{\tt lorenz/}. 
\bibitem{Maju96a} S.N. Majumdar, A.J. Bray, S.J. Cornell and C. Sire, 
Phys. Rev. Lett. {\bf 77}, 3704 (1996).
\bibitem{Maye04} P. Mayer, PhD thesis, King's college London (2004).
\bibitem{Maye05a} P. Mayer and P. Sollich, Phys. Rev. {\bf E71}, 046113 (2005). 
\bibitem{Maye06} P. Mayer, S. L\'eonard, L. Berthier, J.P. Garrahan and P.
Sollich, Phys. Rev. Lett. {\bf 96}, 030602 (2006).
\bibitem{Maze04} G.F. Mazenko, Phys. Rev. {\bf E69}, 016114 (2004).
\bibitem{Meny94} N. Menyh\'ard, J. Phys. {\bf A27}, 6139 (1994).
\bibitem{Murr93} J.D. Murray, {\it Mathematical biology}, 2$^{\rm nd}$ edition,
Springer (Heidelberg 1993). 
\bibitem{Newm90} T.J. Newman and A.J. Bray, J. Phys. {\bf A23}, 4491 (1990). 
\bibitem{Nied72} U. Niederer, Helv. Phys. Acta {\bf 45}, 802 (1972).
\bibitem{Odor06} G. \'Odor, {\tt cond-mat/0606724}. 
\bibitem{Osha89a} G.S. Oshanin and S.F. Burlatsky, J. Phys. {\bf A22}, 
L973 (1989). 
\bibitem{Osha89b} G.S. Oshanin, A.A. Ovchinnikov and S.F. Burlatsky, J. Phys. 
{\bf A22}, L977 (1989). 
\bibitem{Paes04} M. Paessens and G.M. Sch{\"u}tz, J. Phys. A: Math. Gen. 
{\bf 37}, 4709 (2004). 
\bibitem{Paul04} R. Paul, S. Puri and H. Rieger, Europhys. Lett. {\bf 68},
881 (2004).
\bibitem{Paul05} R. Paul, S. Puri and H. Rieger, Phys. Rev. {\bf E71}, 061109
(2005).
\bibitem{Perr77} M. Perroud, Helv. Phys. Acta {\bf 50}, 233 (1977).  
\bibitem{Pico02} A. Picone and M. Henkel, J. Phys. {\bf A35}, 5575 (2002).
\bibitem{Pico04} A. Picone and M. Henkel, Nucl. Phys. {\bf B688}, 217 (2004).
\bibitem{Plei03} M. Pleimling, {\tt private communication}. 
\bibitem{Plei04} M. Pleimling, Phys. Rev. {\bf B70}, 104401 (2004). 
\bibitem{Plei04a} M. Pleimling and F. Igl\'oi, Phys. Rev. Lett.
{\bf 92}, 145701 (2004). 
\bibitem{Plei05} M. Pleimling and A. Gambassi, Phys. Rev. {\bf B71}, 
180401(R) (2005).
\bibitem{Rama04} J.J. Ramasco, M. Henkel, M.A. Santos and C.A. de Silva 
Santos, J. Phys.  {\bf A37}, 10497 (2004).
\bibitem{Roet06} A. R\"othlein, F. Baumann and M. Pleimlimg, Phys. Rev. {\bf E}
at the press {\tt cond-mat/0609707}.
\bibitem{Sant97} J.E. Santos, J. Phys. {\bf A30}, 3249 (1997). 
\bibitem{Sast03} F. Sastre, I. Dornic and H. Chat\'e, Phys. Rev. Lett.
{\bf 91}, 267205 (2003). 
\bibitem{Schm95} B. Schmittmann and R.K.P. Zia, in C. Domb and J. Lebowitz (eds)
{\it Phase transitions and critical phenomena}, Vol. 17, London (Academic 1995).
\bibitem{Sche03} G. Schehr and P. Le Doussal, Phys. Rev. {\bf E68}, 046101 (2003). 
\bibitem{Sche05} G. Schehr and R. Paul, Phys. Rev. {\bf E72}, 016105 (2005).
\bibitem{Sche06} G. Schehr and R. Paul, J. Phys. Conf. Series {\bf 40}, 27 (2006). 
\bibitem{Schu00} G.M. Sch{\"u}tz, in C. Domb and J. Lebowitz (eds) {\sl Phase 
Transitions and Critical Phenomena}, Vol. 19, London (Academic 2000), p.1.
\bibitem{Sigg77} E. Siggia, Phys. Rev. {\bf B16}, 2319 (1977). 
\bibitem{Stoi05} S. Stoimenov and M. Henkel, Nucl. Phys. {\bf B723}, 205 (2005).
\bibitem{Stru78} L.C.E. Struik, {\it Physical ageing in amorphous polymers and
other materials}, Elsevier (Amsterdam 1978).
\bibitem{Taeu05} U.C. T\"auber, M. Howard and B.P. Vollmayr-Lee, 
J. Phys. A: Math. Gen. {\bf 38}, R79 (2005).
\bibitem{Taeu06} U.C. T\"auber, in \cite{Henk06} ({\tt cond-mat/0511743}).
\bibitem{vKam92} N.G. van Kampen, {\it Stochastic porcesses in physics and 
chemistry}, 2$^{\rm nd}$ edition, North Holland (Amsterdam 1992).
\bibitem{Voit05a} R. Voituriez, M. Moreau and G. Oshanin, Europhys. Lett. 
{\bf 69}, 177 (2005). 
\bibitem{Voit05b} R. Voituriez, M. Moreau and G. Oshanin, J. Chem. Phys. {\bf 122}, 084103 (2005). 
\bibitem{Wijl98} F. van Wijland, R. Oerding and H.J. Hilhorst, Physica
{\bf A251}, 179 (1998).
\bibitem{Wolf90} D.E. Wolf and J. Villain, Europhys. Lett. {\bf 13}, 389 (1990).
\bibitem{Ziff86} R. Ziff, E. Gulari and Y. Barshad, Phys. Rev. Lett. {\bf 56}, 
2553 (1986). 
\bibitem{Zwan01} R. Zwanzig, {\it Nonequilibrium statistical mechanics},
Oxford University Press (Oxford 2001). 







\end{thebibliography}
\end{document}